\documentclass[letterpaper,english,twocolumn, floatfix, aps,amsfonts,amssymb, prb,superscriptaddress,showpacs]{revtex4-1}

\usepackage[english]{babel}
\usepackage[applemac]{inputenc}
\usepackage{pgf}
\usepackage{amsmath}
\usepackage{bbold}
\usepackage{graphicx}
\usepackage{dsfont}
\usepackage{bm}

\usepackage{tikz}
\usepackage{units}
\usepackage{dsfont}
\usepackage[amssymb]{SIunits}
\usetikzlibrary{decorations.pathmorphing}
\usepackage{graphicx}
\usepackage{delarray}

\newcommand{\ie}{i.~e.,} 

\newcommand{\hc}{{\text{H.\ c.}}}  
\newcommand{\ket}[1]{\mid #1\rangle}

\newcommand{\bracket}[3]{\langle#1|\,#2\,|#3\rangle}

\newcommand{\vv}{V}

\newcommand{\drg}[1]{\mathcal{D}_{#1}}

\newcommand{\blank}[1]{}

\newcommand{\amd}[1]{a_{#1}^{\dagger}}
\newcommand{\am}[1]{a_{#1}^{\phantom{\dagger}}}

\newcommand{\ampm}{a_{m\pm}^{\phantom{\dagger}}}

\newcommand{\f}[1]{f_{#1}^{\phantom{\dagger}}}
\newcommand{\fd}[1]{f_{#1}^{\dagger}}

\newcommand{\rhs}{right-hand side}

\newcommand{\nr}{\nonumber\\}

\newcommand{\gdop}[1]{g_{#1}^{\dagger}}
\newcommand{\gop}[1]{g_{#1}^{\phantom{\dagger}}}

\newcommand{\trg}[1]{\mathcal{T}[#1]}
\newcommand{\phidop}[1]{\phi_{#1}^{\dagger}}
\newcommand{\phiop}[1]{\phi_{#1}^{\phantom\dagger}}
\newcommand{\oop}[2]{\mathcal{O}_{#2}^{#1}}

\newcommand{\hzero}{H_{N,0}}
\newcommand{\ww}{W}

\begin{document}
\title{Kondo dynamics in one-dimensional doped ferromagnetic
  insulators} 
\date{\today}

\author{Hudson Pimenta}
\affiliation{Department of Physics, University of Toronto, 60
  St. George Street, Toronto, Ontario M5S 1A7, Canada}
\affiliation{Instituto de F\'{i}sica de S\~ao Carlos, Universidade de S\~ao Paulo, C.P. 369, S\~ao Carlos, SP,  13560-970, Brazil}

\author{Luiz N. Oliveira}
\affiliation{Instituto de F\'{i}sica de S\~ao Carlos, Universidade de S\~ao Paulo, C.P. 369, S\~ao Carlos, SP,  13560-970, Brazil}

\author{Rodrigo G. Pereira}
\affiliation{Instituto de F\'{i}sica de S\~ao Carlos, Universidade de S\~ao Paulo, C.P. 369, S\~ao Carlos, SP,  13560-970, Brazil}

\begin{abstract}

Some well-established  examples of itinerant-electron  ferromagnetism in one dimension   occur in a Mott-insulating phase. We examine the   consequences of  doping a ferromagnetic insulator and coupling magnons to gapless charge fluctuations. Using a bosonization scheme for strongly interacting electrons, we derive an effective field theory for the magnon-holon interaction. When the magnon momentum matches the Fermi momentum of the holons, the backscattering of the magnon at low energies  gives rise to a Kondo effect of a pseudospin defined from the chirality degree of freedom (right- or left-moving particles). The crossover between weak-coupling and strong-coupling fixed points  of the effective mobile-impurity model is then  investigated using a numerical renormalization group approach. 

\end{abstract}

\maketitle

\section{Introduction}
\label{sec:introduction}
Ferromagnetism remains a challenging problem in physics, despite having been
investigated ever since (and even before) the advent of quantum mechanics. It was
Heisenberg who first realized that the phenomenon results from an
interplay between electron-electron interactions and the Pauli exclusion principle: 
\cite{Heisenberg:1928tv} when two spin-polarized electrons   occupy two
orthogonal orbitals, their wave function must be spatially antisymmetric and vanishes when they occupy the same position. This leads 
to a lower expectation value of the Coulomb repulsion for electrons with the same spin.  The  dependence of   energy levels on the relative spin orientation is often cast as an effective exchange interaction, which in this case is ferromagnetic as it favors parallel alignment of spins. 


The mechanism of direct exchange  can be generalized to the many-electron problem.  However, its  implications for the ground state are not so clear. 
On the one hand, mean-field   arguments predict that a gas of itinerant electrons with 
local repulsive interactions will spontaneously break spin rotational symmetry and become
ferromagnetic for sufficiently strong interaction. The precise
condition for the transition is given by the Stoner criterion: $U \rho (E_F) > 1$, where
$U$ is the interaction strength and $\rho(E_F)$ is the density of states
at the Fermi level. \cite{Fazekas:1999ud}

On the other hand, the Stoner criterion is not entirely  reliable
because the putative transition occurs in a nonperturbative regime of large
$U$.  Experimentally, the criterion remains controversial. In transition metal oxides, the existence of a ferromagnetic phase depends on   detailed information about the band structure.\cite{Fazekas:1999ud} Recent  observations in cold-atom systems have concluded first  in favor of and later against an interaction-driven 
ferromagnetic transition. \cite{Jo:2009uu, Sanner:2012gf}

Itinerant ferromagnetism has proved hard to establish in microscopic models beyond mean-field approximations. For many
years,   explicit proofs    relied on
peculiar conditions such as   the limit of vanishing hole doping
\cite{Nagaoka:1966cm} or the presence of flat
bands. \cite{Mielke:1991hx} In the domain
of one-dimensional systems,    there are even more constraints: a theorem due to Lieb and Mattis\cite{Lieb:1962wa}  rules out a
ferromagnetic ground state for a number of models,
including the paradigmatic   Hubbard model.

The Lieb-Mattis theorem does not hold  for models with hopping beyond nearest neighbors. In fact, Tasaki\cite{Tasaki:1995do}  proposed a two-band one-dimensional model whose ground state  can  be shown  to be  fully polarized for a
wide range of hopping parameters and finite repulsion. Remarkably, the proof relies on the condition of
quarter-filling, for which the model was conjectured to be a Mott
insulator. One might then wonder whether ferromagnetic order survives in the metallic phase, reached by electron or hole doping.
It is believed that it does, but
the evidence relies mostly on variational methods and exact diagonalization
 for small chains.\cite{Penc:1996td,Fazekas:1999ud} There is stronger evidence
based on density matrix renormalization group (DMRG) results  for the single-band Hubbard model with next-nearest-neighbor hopping. \cite{Daul:1998jba} Only in the infinite-repulsion limit has metallic ferromagnetism been rigorously established, for a multi-band model with no flat bands.\cite{Tanaka:2007ho}

The nature of the ferromagnetic  transition in one dimension has also been studied. Paramagnetic one-dimensional
metals  behave as Luttinger liquids, in which charge and spin excitations are
described by two independent  charge and spin  bosonic fields.
\cite{Haldane:1981vf} As the interaction increases and transition to a ferromagnetic phase supposedly  
occurs, the spin velocity must become negative.
\cite{Daul:1998jba, Yang:2004kv, Wang:2005wf} Theories have
been proposed  to describe second-order transitions for
Ising \cite{Yang:2004kv,Bartosch:2003fy}  and $SU(2)$ \cite{Sengupta:2005kra} symmetries. First-order transitions are also a possibility.
\cite{Nishimoto:2008gi, Takayoshi:2010bn}

In this work, we investigate the stability of  doped
Mott-insulating ferromagnets in one dimension at zero temperature  by considering the
creation of a magnon and its interaction with gapless charge fluctuations. This approach  follows the spirit of
current experiments designed to investigate phase transitions in cold atomic gases. 
\cite{Koschorreck:2012fj, Fukuhara:2013hq}

We organize our presentation as follows. Our starting point in
Sec.~\ref{sec:models-with-two} is a
generalized version of the models proposed in Refs.~\onlinecite{Tasaki:1995do}
and \onlinecite{Penc:1996td}. In the weakly
interacting regime, we bosonize the model and show that the charge sector is
a Mott insulator for quarter-filling while the Luttinger liquid   in the  spin sector remains
stable. Section ~\ref{sec:strongly-inter-regim}
considers the model for the ferromagnetic phase   in the strongly interacting regime. We  use an alternative bosonization scheme 
\cite{Matveev:2007ei,Akhanjee:2007ca} to identify the spin and charge
excitations in the large-$U$ limit  as magnons and fermionic holons,
respectively. In Sec.~\ref{sec:effect-field-theory} we analyze the low-doping limit of the metallic phase using an effective field theory with 
 magnon-holon interactions  dictated by symmetry considerations. A similar approach has been used to study 
magnons in a spinor Bose liquid. \cite{Kamenev:2009ez} We find that,
when the magnon momentum is commensurate with the holon Fermi surface, the scattering between opposite Fermi points gives rise to  infrared singularities 
akin to the Kondo effect.\cite{Hewson:1997vc} This phenomenon  has been previously discussed in the context of a mobile impurity in a Luttinger liquid.\cite{Lamacraft:2009bp,Schecter2012639} In Sec. \ref{sec:1}, we   proceed to  studying the low-energy fixed points of our effective model using the numerical renormalization group.\cite{KWW80:1003,BCP:2007}  Finally, Section~\ref{sec:summary} summarizes our results.

\section{ Weakly Interacting Regime}
\label{sec:models-with-two}
In Sec. \ref{sec:introduction}, we briefly mentioned   Tasaki's model, a
one-dimensional model for which ferromagnetism has been rigorously
established. \cite{Tasaki:1995do} 
Tasaki's Hamiltonian is
\begin{equation}
  \label{tasakiHamiltonian}
  H = \sum_{i, j,\sigma} t_{ij} c^\dagger_{i,\sigma} c_{j,\sigma} + U \sum_i n_{i, \uparrow}
n_{i, \downarrow},
\end{equation}
where $c_{i,\sigma}$ is the annihilation operator for an electron with  spin $\sigma=\uparrow,\downarrow$ at site $i$, $n_{i,\sigma}=c^\dagger_{i,\sigma}c^{\phantom\dagger}_{i,\sigma}$ is the number operator, $U\geq0$ is the on-site repulsion, and  the hopping amplitudes  $t_{ij}$ are defined as follows:
\begin{align*}
t_{i, i+1} &= t',& & \forall\, i,\nonumber \\
t_{i, i+2} &= t ,  & &  \text{if $i$ is even}, \nonumber \\
t_{i,i+2} &= -s,   & &  \text{if $i$ is odd}, \nonumber \\
t_{i,i} &= V,   & &  \text{if $i$ is odd}, \nonumber \\
t_{i,j} &= 0,   & &  \text{otherwise.} \nonumber 
\end{align*}

\begin{figure}[t]
\includegraphics[width=.75\columnwidth]{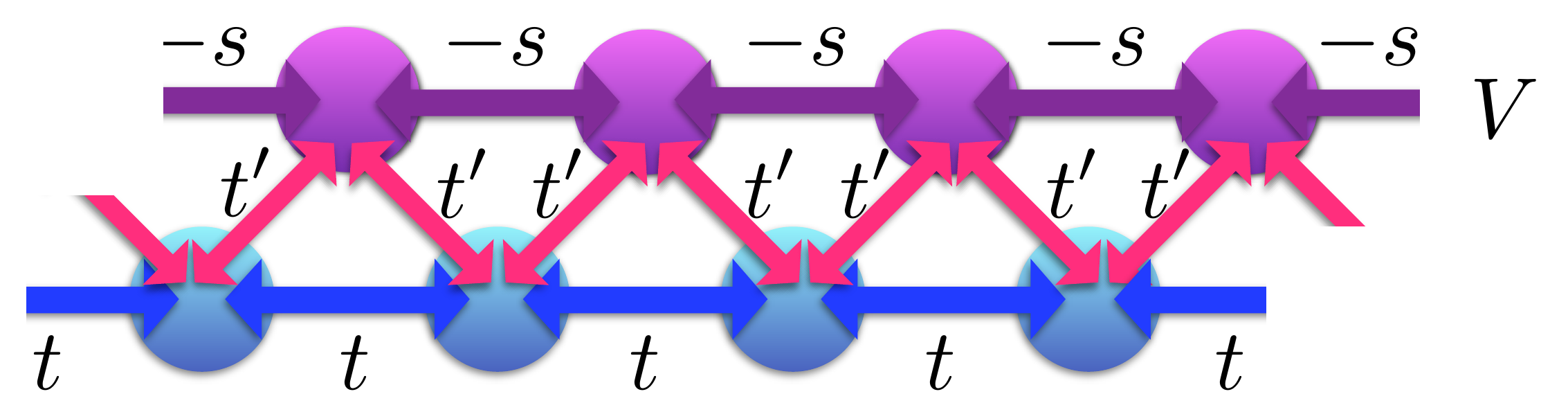}
\caption[tasakilattice]{\label{tasakilattice} (color online) Tasaki's one-dimensional model. The hopping between nearest neighbors is
$t'$. The hopping
between next-nearest neighbors is $t$ ($-s$) for the even  (odd) sublattice.  The odd sublattice has chemical potential $V$. The
on-site electronic repulsion in both sublattices  is $U$.  }
\end{figure}

Fig.~\ref{tasakilattice}
illustrates the Tasaki lattice. For $t\neq-s$, the lattice has two sites per unit cell. 
Tasaki demonstrated
 the existence of a ferromagnetic phase at 
quarter-filling when $t$
and $U$ are sufficiently large compared to $s$  and for the particular choices $V = 0$ and  $t' =
\sqrt{2} \left( t + s \right)$. Under these conditions, it can be shown
that the expectation value of the Hamiltonian in an arbitrary state  
satisfies the   inequality $\langle H
\rangle \ge - 2 \left(s + t \right) N$, where $2N$ is the total number of
sites. The lower bound  happens to be precisely the energy of the fully
polarized   state, which proves that the latter is one of the ground states.\cite{Tasaki:1995do} The ground state is  $(2S_{tot}+1)$-fold degenerate, with $S_{tot}=N/2$ at quarter-filling, due to spontaneous symmetry breaking of the $SU(2)$ symmetry.

Let us  describe the low-energy excitations of Tasaki's model, starting with the paramagnetic phase
at weak coupling. As a first step, we   diagonalize the Hamiltonian for $U = 0$, which we denote by
$H_0$. Using the momentum representation defined on the even and the odd sublattices 
\begin{eqnarray}
  \label{c2j}
 c_{2j, \sigma} &=& \frac{1}{\sqrt{N}}\sum_k e^{ik (2j)} a_{k, \sigma}, \\
\label{c2j+1}
c_{2j+1, \sigma} &=& \frac{1}{\sqrt{N}} \sum_k e^{ik (2j+1)}  b_{k,\sigma},  
\end{eqnarray}
with $ k\in [-\pi/2, \pi/2]$ in the first Brillouin zone, we can rewrite $H_0$  as
\begin{eqnarray}
H_0 &=& \displaystyle\sum_{k,\sigma} \left[  2t \cos (2k) a_{k,\sigma}^\dagger
  a_{k,\sigma} - 2s \cos (2k) b_{k,\sigma}^\dagger b_{k,\sigma} \right.\nonumber
  \\
&& \left.   + 2t' \cos k ( b_{k,\sigma}^\dagger a_{k,\sigma} +  \hc )  \right].
\end{eqnarray}
We see now that   Tasaki's  model is a particular case of a model of
the form
\begin{eqnarray}
\label{modelK}
H_0 &=& \displaystyle\sum_{k,\sigma} \left[ \epsilon_a (k)  a_{k,\sigma}^\dagger
  a_{k,\sigma} + \epsilon_b (k)  b_{k,\sigma}^\dagger b_{k,\sigma} \right.\nonumber
  \\
&& \left.   + \alpha(k) ( b_{k,\sigma}^\dagger a_{k,\sigma} +  h. c.  )  \right].
\end{eqnarray}
This model can be diagonalized through a rotation \begin{eqnarray}
\label{bk}
a_{k,\sigma} =\cos \left( \frac{\theta_k}{2} \right) g_{+,k,\sigma} -\sin \left( \frac{\theta_k}{2} \right) g_{-,k,\sigma},\\
b_{k,\sigma} =\sin \left( \frac{\theta_k}{2}\right) g_{+,k,\sigma}+ \cos \left( \frac{\theta_k}{2} \right) g_{-, k,\sigma},
\label{ck}
\end{eqnarray}
where
\begin{equation}
\label{anglerotation}
\tan \theta_k \equiv \frac{ 2\alpha(k) } { \epsilon_a (k) -
    \epsilon_b (k) }.
\end{equation}
We obtain  two bands
\begin{equation}
  \label{eq:2}
  H_0 = \sum_{k, \sigma} \left[ \epsilon_+ (k) g_{+, k, \sigma}^\dagger g_{+, k,
    \sigma} + \epsilon_-(k) g_{-, k, \sigma}^\dagger g_{-, k,
    \sigma} \right],
\end{equation}
with dispersion relations 
\begin{equation}
\epsilon_{\pm} (k) = \displaystyle\frac{\epsilon_a (k) + \epsilon_b (k) }{2}
\pm \sqrt{
  \alpha^2 (k) + \left[ \frac{\epsilon_a (k)  - \epsilon_b (k)}{2} \right]^2}.
\end{equation}
The  noninteracting ground state is paramagnetic.  At quarter-filling the single-particle states in the lower band  with   $|k| \le
k_F = \pi/4$ are occupied   (corresponding to half-filling of the lower band). The neutral excitations are electron-hole pairs on top of this ground state.

Now we consider the weakly interacting regime  $U\ll v_F$,
where $v_F = \left. \frac{d\epsilon_- } {dk}
\right|_{k = k_F}$ is the   Fermi velocity. The weak-coupling  condition justifies linearizing the low-energy spectrum 
around $\pm k_F$. Through bosonization,\cite{vonDelft:1998ff,Miranda:2003wz} one can map the linearized version
of the kinetic energy $H_0$ onto an effective Hamiltonian
\begin{equation}
  \label{stringHamiltonian}
  H_{LL}^{0} = \sum_{\nu = c, s} \int dx~ \frac{v_F}{2} \left [ 
    \Pi_\nu^2 +  \left( \partial_x \phi_\nu\right)^2
  \right],
\end{equation}
where $\phi_c$ and $\Pi_c$  ($\phi_s$ and $\Pi_s$) are the charge
(spin) canonically-conjugated bosonic fields. It can be shown that $\partial_x \phi_c(x) \sim \rho(x)$
and $\partial_x \phi_s(x) \sim m(x)$, where $\rho(x)$ is the charge density and
$m(x)$ is the local magnetization. The remarkable feature of bosonization
is that, once the interacting term is bosonized as well, charge and
spin fields remain noninteracting, the Hamiltonian of
Eq.~\eqref{stringHamiltonian} being only slightly modified:\cite{Miranda:2003wz}
\begin{equation}
  \label{luttingerliquid}
  H_{LL} = \sum_{\nu = c, s} \int dx~ \frac{v_\nu}{2} \left [ K_\nu
    \Pi_\nu^2 + \frac{1}{K_\nu} \left( \partial_x \phi_\nu\right)^2
  \right].
\end{equation}
Hamiltonian $H_{LL}$ is known as the Luttinger model; $v_c$ and $v_s$ are the
charge and spin velocities; and $K_c$ and $K_s$ are the Luttinger
parameters in the charge and spin sectors, respectively. In the noninteracting
case, $v_c = v_s = v_F$ and $K_c = K_s = 1$.   For repulsive interactions, we have $K_c
< 1$, while spin  SU(2) symmetry fixes    $K_s = 1$.\cite{Giamarchi:2004uc}
 The   interaction has the effect of   modifying the velocities and Luttinger
parameters, leading to the phenomenon of spin-charge separation: in the low-energy limit, the elementary excitations are  independent  charge and spin collective modes.  

At weak coupling, we can determine the $U$ dependence of charge and spin velocities by directly bosonizing the interaction term in Eq. (\ref{tasakiHamiltonian}). For this purpose, we first  promote the site operators to fields. Defining a field is
not straightforward as the expressions for $c_{2j, \sigma}$ and
$c_{2j+1, \sigma}$ are different. Combining Eqs.~(\ref{c2j}),
(\ref{c2j+1}), (\ref{bk}) and (\ref{ck}) to express the site operators
in terms of $g_{\pm,k,\sigma}$  yields
\begin{eqnarray}
   \label{c2jg}
 c_{2j, \sigma} \sim - \sum_k \frac{e^{ik (2j)}}{\sqrt{N}}   \sin \left(
   \frac{\theta_k}{2} \right) g_{-,k,\sigma} + \cdots, \\
\label{c2j+1g}
c_{2j+1, \sigma} \sim \sum_k \frac{e^{ik (2j+1)}}{\sqrt{N}} \cos \left(
  \frac{\theta_k}{2} \right) g_{-, k,\sigma} + \cdots,  
\end{eqnarray}
where we have omitted terms proportional to $g_{+,k,\sigma}$ because
the latter  involve high-energy excitations, in the upper band.

We can combine Eqs.~\eqref{c2jg} and \eqref{c2j+1g} into a single field  as follows
\begin{equation}
  \label{psifield}
  \psi_\sigma (x) \sim \sum_k e^{ i k x} f(k, x) g_{-, k, \sigma},
\end{equation}
where we have introduced the function \begin{equation}\label{falternates}
f(k,x)=\frac{1+e^{i\pi x}}{2}\cos\left(\frac{\theta_k}{2}\right)-\frac{1-e^{i\pi x}}{2}\sin\left(\frac{\theta_k}{2}\right),
\end{equation}
which reduces to $f(k,x)=\cos(\theta_{k}/2)$ [$f(k,x)=\sin(\theta_{k}/2)$] for the positions $x$ belonging to the
even (odd) sublattice.

The next step is to expand  the electron  field into 
right and left movers
\begin{equation}
  \label{linearizedpsi}
  \psi_\sigma(x) \sim  f(k_F, x) \left[ e^{-i k_F x} \psi_{L,\sigma} (x) + e^{i k_F x}
    \psi_{R,\sigma} (x) \right],
\end{equation}
where
\begin{equation}
  \label{psirightandleft}
  \psi_{(R,L), \sigma} (x) \sim\frac1{\sqrt{N}} \sum_{p=-\infty}^{+\infty} e^{\pm i p x} g_{(R,L),p,\sigma}.
\end{equation}
Here $g_{(R,L), p, \sigma}$ are the annihilation operators of the
linearized branches around the right and left Fermi points.  

In the continuum limit, the interaction term in Eq. (\ref{tasakiHamiltonian}) becomes $U\sum_{j}n_{j,\uparrow}n_{j,\downarrow}\approx  U\int dx\, \psi^\dagger_{\uparrow}\psi^{\phantom\dagger}_{\uparrow} \psi^\dagger_{\downarrow}\psi^{\phantom\dagger}_{\downarrow}$. The procedure to bosonize the interaction    is now almost identical with that for the Hubbard model,\cite{Giamarchi:2004uc}  except for the 
 alternation between even and odd sublattices that  introduces the factor of $  f(k_F,x)$.  
From now on, we follow  the same steps
as for the Hubbard model, but must be careful to take into account  the oscillations of  $f(k_F, x)$.

The uniform part of the density operator, combined with the $x$-independent  part of  $f(k_F, x)$,  gives rise to terms in the effective Hamiltonian which are quadratic in the bosonic fields, as in Eq. (\ref{luttingerliquid}).  Compared with the result for the Hubbard model, the effective $U$ is renormalized into an effective  $\tilde{U} =\frac{1}{4} \left[ 5+ \cos (2
  \theta_{k_F})\right] U$. (The proportionality factor actually depends
on the choice of $f(k, x)$ introduced in Eq.~\eqref{falternates}, but
the important result is that $\tilde{U} \sim U$.) Therefore, one must just replace $U$ by $\tilde{U}$ in the expressions
for the Luttinger parameters of the Hubbard model.
 In particular, the spin velocity for $U\ll v_F$ is known
to be \cite{Giamarchi:2004uc}
\begin{equation}
  \label{spinonvelocity}
  v_s \approx  v_F \left( 1 - \frac{U}{\pi v_F} \right)^{1/2} \to v_F \left( 1 - \frac{\tilde{U}}{\pi v_F} \right)^{1/2}.
\end{equation}

The Luttinger liquid phase becomes unstable
when $v_s \to 0$, which we can interpret as  a sign of a  phase transition to a state with spontaneous magnetization.\cite{Yang:2004kv}  In principle, one could search for this instability
by extrapolating the result in  Eq.~(\ref{spinonvelocity}). However, the condition $\tilde U = \pi
v_F$ is not compatible with $U \ll v_F$, required for  weak-coupling 
bosonization to  hold.   In the next section we shall consider the effective field theory in the   strong-coupling limit.  Before doing so, we now argue that the model
is  a Mott insulator at quarter-filling, even at arbitrarily small $U$.

The insulating behavior is due to Umklapp scattering, which becomes commensurate in the two-sublattice system when $k_F=\pi/4$. The Umklapp operator stems from the interaction term  $\sim U\int dx [f(k_F,x)]^4e^{i4k_Fx}\psi^\dagger_{L,\uparrow} \psi^{\phantom\dagger}_{R,\uparrow}\psi^\dagger_{L,\downarrow}\psi^{\phantom\dagger}_{R,\downarrow}+\hc$. 
The  oscillating
component of $f(k_F, x)$ can be  combined with the  
oscillations of the fermionic field  to generate terms proportional to $e^{i \left(\pi \pm 4
  k_F\right) x}$. Often one argues that these terms oscillate rapidly and can be
neglected in the low-energy Hamiltonian. However, precisely for quarter-filling, the oscillations
cease and these terms must be kept. The bosonized version of the 
nonoscillating Umklapp process is 
\begin{equation}
  \label{sinegordonextraterm}
    H_{Umklapp} = g \int dx~  \cos \left( { \sqrt{8
      \pi} \phi_c} \right),
\end{equation}
where $g$ is the coupling constant. From the renormalization-group analysis of the sine-Gordon model,\cite{Coleman:1975tn,gogolin2004bosonization}  it is known that this operator has scaling dimension $2K_c$ and   is relevant for arbitrarily weak repulsive interactions. Its effect  is to open up a gap $\Delta_c$ in the charge sector, which for small $g$ scales as $\Delta_c \sim |g|^{1/(2-2K_c)}$.  

A particularly simple result is obtained at  the Luther-Emery point  $K_c = 1/2$, in which the sine-Gordon model can be refermionized into noninteracting spinless fermions.\cite{Luther:1974iz, Schulz:1980wr} The gap can be seen explicitly in the massive relativistic  dispersion  
\begin{equation}
  \label{fermionsdiracdispersion}
  \epsilon_{c,\pm}(p) = \pm \sqrt{ (v_cp)^2 + \Delta_c^2 }. 
\end{equation} 
The  positive (negative) energy, as measured from the chemical potential, refers to a completely empty (filled) band of free fermions which are the elementary excitations in the charge sector. The gap between the bands at the Luther-Emery point is $2\Delta_c\sim |g|$. 

In our case, a caveat is necessary. Setting $k_F=\pi/4$ in Eq. (\ref{falternates}) yields $g=0$ since $\theta_{k=\pi/4}=\frac{\pi}{2}\textrm{sgn}(t^\prime)$ for Tasaki's model (see Eq. (\ref{anglerotation})). However, this bosonization procedure only predicts the  coupling constants correctly to first order  in $U$. Since the Umklapp operator is not forbidden by any symmetries in Tasaki's model, we expect it to be generated at higher order, most likely  $g\sim \mathcal O(U^2)$. Therefore, the effective field theory predicts the system to be paramagnetic at weak coupling, i.~e., a Luttinger liquid in the spin sector, and to become a charge insulator at quarter filling.

\section{ Strongly interacting regime}
\label{sec:strongly-inter-regim}

We have seen in Sec. \ref{sec:models-with-two} that there is no ferromagnetic
transition for the model of Eq.~\eqref{modelK} at small
$U$. On the other hand, at least for Tasaki's model at  quarter-filling  we know that the ground state becomes a fully polarized ferromagnet at sufficiently large $U$.\cite{Tasaki:1995do} Since we expect the charge gap to increase monotonically with $U$, there should be a transition   from a paramagnetic Mott insulator (with gapless spin excitations described by the Luttinger model)  to a ferromagnetic Mott insulator (with gapless magnons due to the broken $SU(2)$ symmetry). 
Without discussing the nature of the transition (whether first or second  order\cite{Yang:2004kv,Sengupta:2005kra,Nishimoto:2008gi, Takayoshi:2010bn}), we now consider the strongly interacting regime and take the existence of a ferromagnetic ground state for granted. We shall start from  the insulating
phase  at  quarter-filling, but also  investigate the consequences of doping into the metallic phase.

The difficulty in treating the problem in the strongly interacting
regime is that standard bosonization is not applicable. Nonetheless,  an
alternative bosonization scheme for strongly interacting electrons has been developed.\cite{Matveev:2007ei} This approach starts from the picture that, in the limit of infinite repulsion,  
electrons cannot move past each other; as a result,   the
spin of each electron is confined and therefore frozen. One then writes down an effective model for spinless fermions (\emph{holons})
in  the charge sector. Including corrections due to forward scattering  at large, finite $U$ 
gives rise to an  exchange interaction $J$, which allows one to treat the spin degrees of freedom as an effective spin chain. In
this scenario, spin-charge separation still
holds.\cite{Matveev:2007ei}

We follow this strategy, but  adapt it to the problem
considered here. The main difference is that, in the presence of  next-nearest-neighbor
hopping, an electron can hop around another electron that occupies  a nearest-neighbor
site. Therefore, the spin is not confined even for infinite repulsion. To apply the picture of Ref. \onlinecite{Matveev:2007ei}, we consider the more general Hamiltonian  in 
Eq.~\eqref{tasakiHamiltonian} with  chemical potential $V \neq 0$.  Recall that Tasaki's proof of a ferromagnetic ground state only applies to the case $V=0$, $t^\prime=\sqrt{2}(t+s)$. However, adding the staggered chemical potential  does not change the symmetry of the Hamiltonian, and it has been argued that the ferromagnetic phase is observed also for $V\neq 0$.\cite{Penc:1996td,Fazekas:1999ud}


Here, we take $V$ to be
negative  and   large, $| V | \gg t^\prime,t,s$, to constrain   the electrons to
move within  the odd sublattice, with hopping amplitude $-s$. Together with a strong on-site repulsion, $U\gg s$, this condition 
suppresses  exchange processes, freezing the electron spin degree of freedom in the limit $|V|,U\to \infty$. The charge
sector can then be described   by spinless fermions, which we
call   holons from now on. In this regime it is easy to see that the system is a Mott insulator at quarter filling, since it corresponds to half-filling of the odd sublattice, with a gap in the charge excitation spectrum  of order $U\gg s$. At the same time, we can think that these gapped holons descend from the fermions for 
the sine-Gordon model at the Luther-Emery point discussed in Sec. \ref{sec:models-with-two}. Extending the
dispersion in Eq. (\ref{fermionsdiracdispersion}) to the large-gap regime and  expanding  for   $|p|\ll \Delta_c\sim U$, we have
\begin{equation}
  \label{fermionicbands}
  \epsilon_{c,\pm} (p) \approx  \pm (\Delta_c + \gamma_{\pm} p^2).
\end{equation}
Here $\gamma_+$ and $\gamma_-$ are parameters inversely proportional to the holon mass in the upper and lower Hubbard bands, respectively. In contrast with Eq. (\ref{fermionsdiracdispersion}), we   
allow for  $\gamma_+\neq \gamma_-$ because the effective model  in the large-$U$ limit  has no Lorentz invariance nor particle-hole
symmetry. In fact, we expect $\gamma_- \sim s$ and $\gamma_+ \sim
t$ from hopping of holes in the odd sublattice and particles in the even sublattice. 

\begin{figure}
\includegraphics[width=0.75\columnwidth]{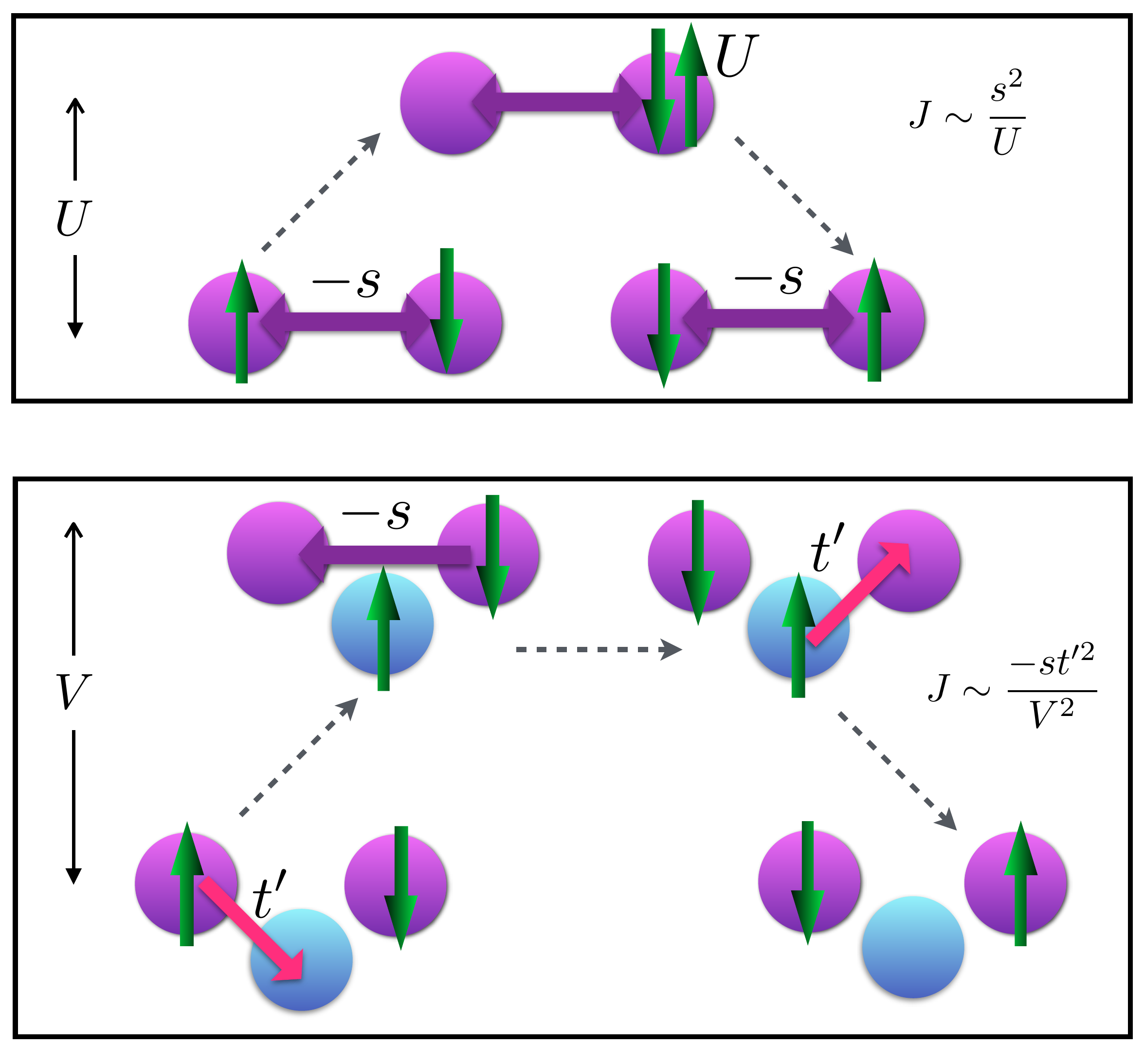}
\caption[Jmechanism]{\label{Jmechanism}  (color online) Spin exchange mechanisms for
  the Tasaki lattice. The first mechanism is the same that takes place in
  the Hubbard model and is always antiferromagnetic. The spin flip
  occurs through an intermediate state of additional energy $U$, as
  represented in the upper box.
 In the second mechanism, represented in the lower box, an
  electron moves to the other sub-lattice, creating a vacancy in the
  first sub-lattice for another electron. Next, the second electron
  takes that vacancy. Finally, the first electron occupy the site of
  the second one. The exchange constant is ferromagnetic
  if  $-s<0$ .}
\end{figure}

Now we take into account that  the repulsion and the chemical potential are finite and
allow spin fluctuations. Treating virtual hopping processes by  perturbation theory leads
to an effective spin exchange interaction, as illustrated in
Fig.~\ref{Jmechanism}. The resulting  Hamiltonian is $H_s
= J \sum_i \mathbf{S}_i \cdot \mathbf{S}_{i+1}$, where $\mathbf{S}_i$ is the spin operator of the $i$-th electron in the chain, and the exchange
  constant is given by \cite{Penc:1996td, Fazekas:1999ud}
\begin{equation}
  \label{Jeffective}
  J = 4\left( \frac{s^2}{U} - \frac{s t'^2}{V^2} \right).
\end{equation}
Note that $J$ can be ferromagnetic only if $s>0$  (recall that  the hopping amplitude in the odd sublattice  is $-s$, hence negative). 
We will take $J < 0$ from now on. In this case, the ground state of the effective spin chain is a fully polarized ferromagnet.

Taking the  ground state to be fully polarized along the $z$ direction, the spin Hamiltonian can   be mapped into magnon excitations
through the Holstein-Primakoff transformation: \cite{Holstein:1940kr}
\begin{eqnarray}
\label{magnetizationandnumberofmagnons}
S_{j}^z &=& S - B_j^\dagger B^{\phantom\dagger}_j, \\
S_{j}^- &=& B_j^\dagger \left( 2S - B_j^\dagger B^{\phantom\dagger}_j
\right)^{1/2} \approx \sqrt{2S} B_j^\dagger, \\
S_j^+ &=& \left( 2S - B_j^\dagger B^{\phantom\dagger}_j \right)^{1/2} B^{\phantom\dagger}_j
\approx \sqrt{2S} B^{\phantom\dagger}_j,
\end{eqnarray}
where $S$ is the spin
quantum number ($S = 1/2$ for electrons), and the magnon operators $B_j$   obey a bosonic
algebra. We see from
Eq.~(\ref{magnetizationandnumberofmagnons}) that the number of magnons
is directly related to the magnetization.

Within standard linear spin-wave theory, the spin Hamiltonian describes noninteracting magnons and  takes the form
\begin{equation}
\label{hdemagnon}
H_s \approx  \sum_k \omega(k) B_k^\dagger B^{\phantom\dagger}_k,
\end{equation}
with $\omega(k) = 2 J S (1 - \cos k) \approx J S k^2$ for $k\ll 1$. The dispersion is gapless because the magnon is the Goldstone mode of the spontaneously broken  $SU(2)$ symmetry. From now on, we
write  simply   $\omega(k) \approx  \lambda k^2$ at low energies.  More generally, the effective $\lambda$ away from the strong-coupling limit can be determined  within a random phase approximation for the spin-spin correlation function.\cite{Mattis:2006ch}  Neglecting   interactions between magnons and gapped charge fluctuations, the ferromagnetic Mott-insulating phase at quarter-filling is stable as long as $\lambda>0$.

\begin{figure}
\includegraphics[width=\columnwidth]{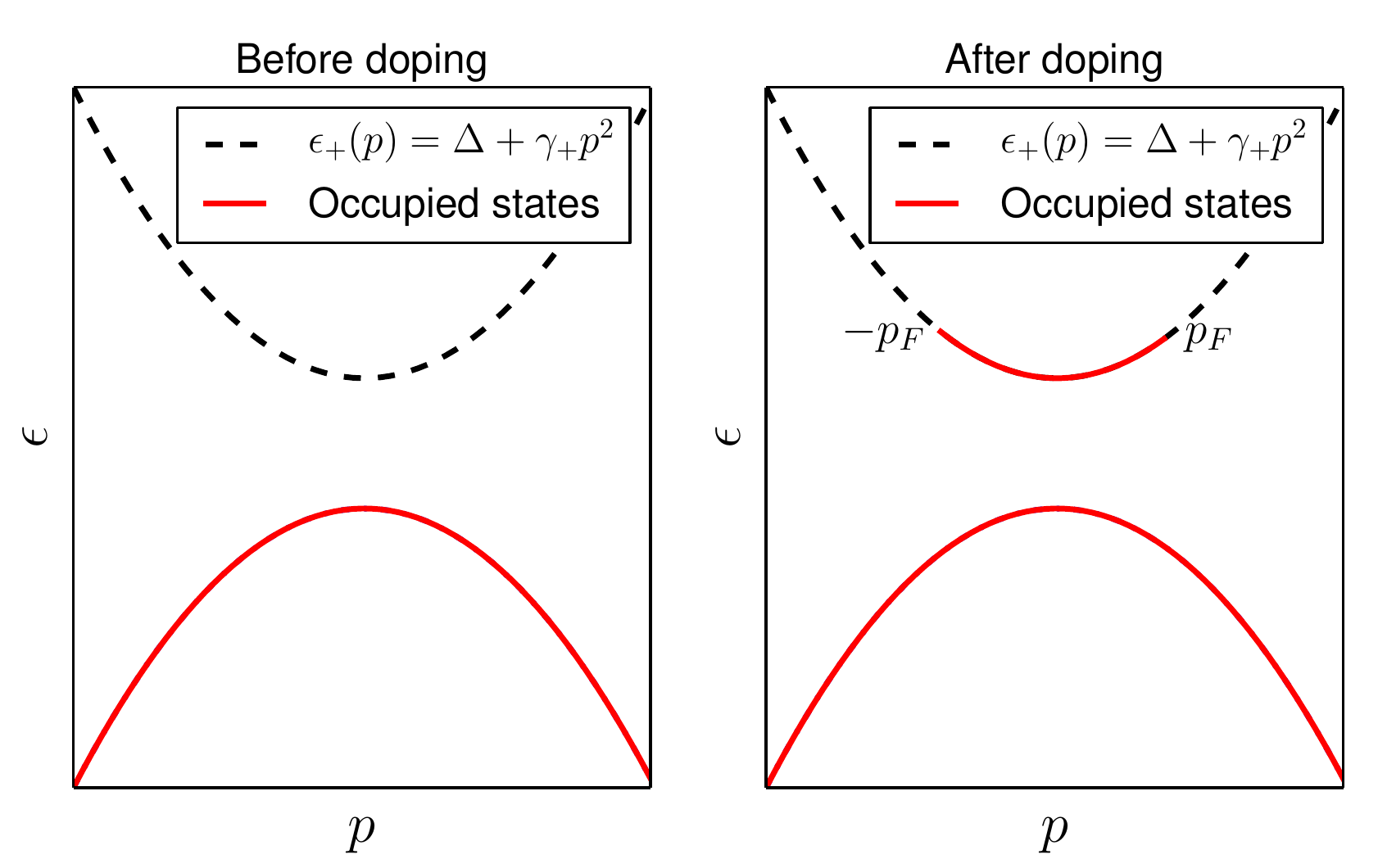}
\caption[fermionband]{\label{fermionband}  (color online) Holon bands  before and after doping the ferromagnetic insulator. At low energies, the
  dispersion is approximately parabolic. The doping introduces two
  Fermi points, $-p_F$ and $p_F$.}
\end{figure}

We are   interested in  the effects of coupling magnons to gapless charge excitations which arise when we move away from quarter-filling. Since the density of holons is directly related to the density
of electrons, adding electrons imply adding holons to the upper Hubbard band. The
configurations before and after doping are illustrated in
Fig.~\ref{fermionband}. The doping  introduces two Fermi points at 
$\pm p_F$, where $p_F$ is related to the    average charge  density $\rho=\sum_{j,\sigma}\langle n_{j,\sigma}\rangle$ by $ p_F=\pi(\rho-1/2)$. At energy scales below the charge gap, we can ignore the lower
band. Moreover, in the low-doping regime (described by the theory of the commensurate-incommensurate transition\cite{Schulz:1980wr}) we can neglect   holon-holon interactions, which are weak even  when $K_c \neq 1/2$.  The charge Hamiltonian is then simply
\begin{eqnarray}
  \label{Hcharge}
  H_c &\approx& \sum_p \epsilon_c(p) c_{p}^\dagger c_{ p},
\end{eqnarray}
   where $c^\dagger_p$ creates holons in the upper  band 
 and $\epsilon_c(p) \approx \gamma p^2$ (omitting the constant  $\Delta_c$ in the energy), with $\gamma \equiv \gamma_+$. (Likewise, removing electrons would create holes in the lower  band. In this case 
one would have to replace $\gamma_+$ by $\gamma_-$ for the low-energy excitations.) 
We recall that in the strongly interacting regime we expect $\gamma\sim t$, whereas $\lambda \sim J$   with $J$ given in Eq. (\ref{Jeffective}).  
Therefore,   our analysis is valid in the regime $\lambda \ll
\gamma$, i.e. when the magnon  mass (in the Galilean sense of a quadratic dispersion) is larger than the holon mass.

We now discuss the interaction
between magnons and holons. The form of the interaction  can be guessed from
symmetry, but before doing so we propose a physical mechanism from which it can
be derived. In general,   the exchange constant $J$  is a function 
  of the average charge  density, $J=J(\rho)$. Following Refs. \onlinecite{Matveev:2004kja,
  Matveev:2004jqa}, we consider that in the metallic phase the exchange interaction  between neighboring spins is a function of the local density:
 $ H_{s} = \sum_i J (\rho_i ) \mathbf{S}_i
\cdot \mathbf{S}_{i+1}$. In our case, $\rho_i$ can fluctuate due to the motion of the dilute electron gas in the even sublattice. Next, we expand $J (\rho_i)$ around the
average density before doping ($\rho_0=1/2$): $J(\rho_i ) \approx J (\rho_0) +
\frac{\delta J}{\delta \rho_i} \big|_{\rho_i = \rho_0} \delta \rho_i$. The term proportional to the derivate
of $J$ leads to the spin-charge interaction
\begin{equation}
  \label{derivativeofJHamiltonian}
  H_{cs} =  \sum_i \frac{\delta J (\rho_i )}{\delta \rho_i}
  \bigg|_{\rho_i = \rho_0} \delta
  \rho_i \mathbf{S}_i
\cdot \mathbf{S}_{i+1}.
\end{equation}
Replacing $\delta \rho_i$ by the holon  density operator of the upper
band, we can write 
\begin{equation}
  \label{derivativeofJHamiltonian2 }
  H_{cs} = -\alpha \sum_i  c^\dagger_i c^{\phantom\dagger}_i \mathbf{S}_i
\cdot \mathbf{S}_{i+1},
\end{equation}
where  $\alpha\equiv - \left.\frac{\partial J}{\partial \rho}\right|_{\rho_0}$ and $c_i$ is the Fourier transform of $c_p$ in Eq. (\ref{Hcharge}). Using  the
Holstein-Primakoff transformation again and
switching to the momentum representation, we obtain the quartic  magnon-holon interaction $H_{I} = \sum_{k,q,p} \zeta(k, p, q)
c_{p-q}^\dagger c_p B_{k+q}^\dagger B_k$, with
\begin{eqnarray}
\zeta (k, p, q) &=&
- \alpha \left[ 
e^{i k} + e^{-i (k+q)} - e^{-i q} - 1 \right] \nonumber \\
 &\approx& \alpha k(k+q).
\end{eqnarray} 
Note that the coupling function is proportional to the momenta of
the incoming and the outgoing magnons. Next we discuss how the same
form is enforced by symmetry.

Due to spin $SU(2)$ symmetry, the total magnetization $S^z_{tot}=\sum_j S_j^z$
is a conserved quantity in our model.  Therefore, according to
Eq.~(\ref{magnetizationandnumberofmagnons}), the number of magnons
must also be conserved. This implies that the magnon-holon interaction vertex $\zeta$
  must contain one incoming and one outgoing magnon legs (in contrast with the more
often encountered coupling between electrons and phonons or
photons). By charge conservation, there is also one incoming and one outgoing holon.

\begin{figure}
\includegraphics[width=0.4\columnwidth]{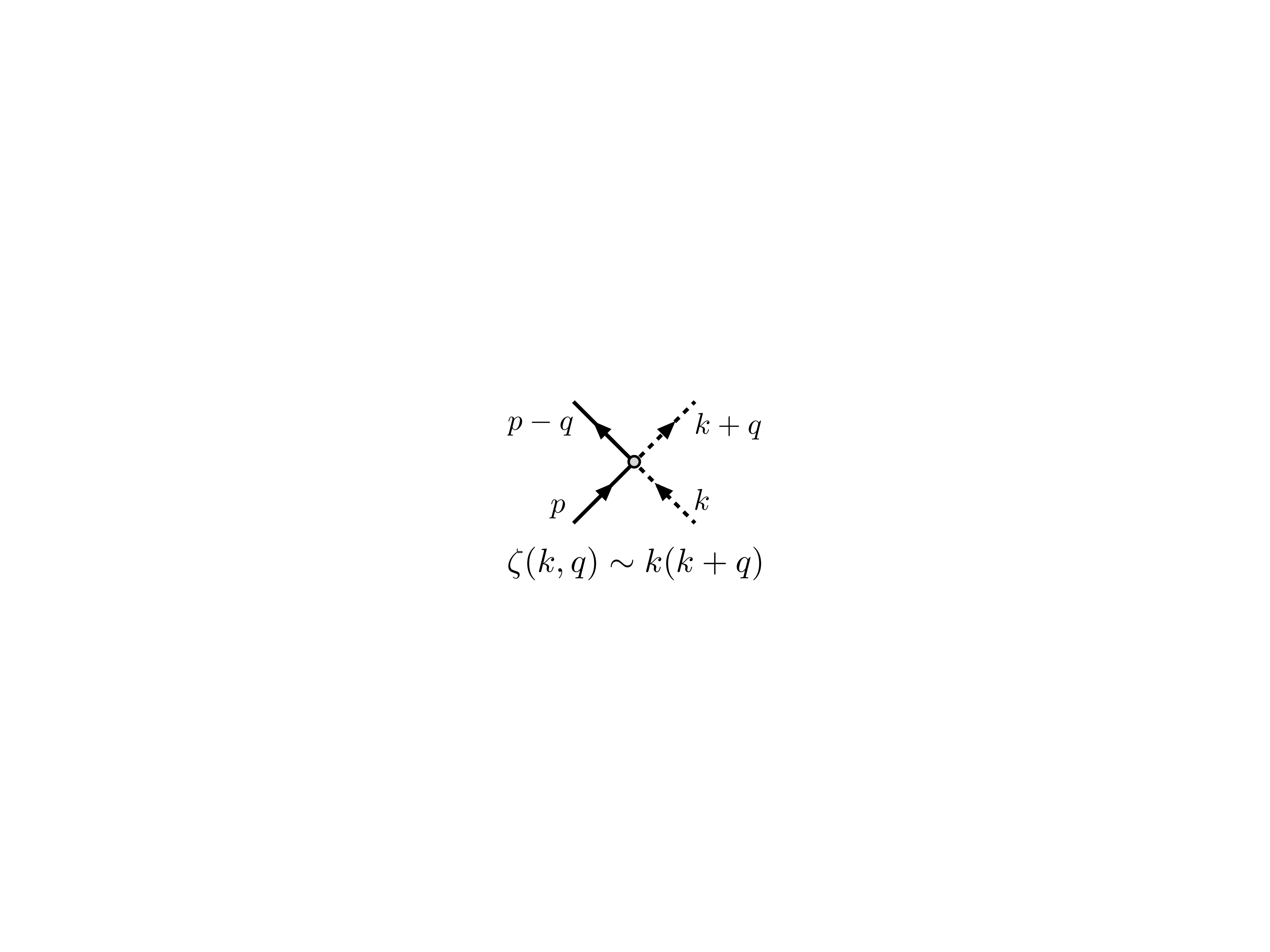}
\caption[couplingdiagram]{\label{couplingdiagram} Magnon-holon
  interaction. Solid (dashed) lines represent holons (magnons).
The momentum dependence of the coupling function
  $\zeta$ is dictated by symmetry.}
\end{figure}

Let $p$ and $k$ be the momenta of the incoming holon and magnon,
respectively, and $p-q$ and $k+q$ the momenta of the outgoing holon
and magnon. In principle, the coupling function $\zeta$ can depend on $p$, $k$
and $q$. Parity symmetry requires that the expansion of $\zeta$ for small $p,k,q$ contains only  even powers of   momenta. Furthermore,  there should be no coupling to magnons with zero momentum (Goldstone bosons) since their presence only amounts to a uniform rotation of the
  magnetization. Thus,  $\zeta \to 0$ if $k \to 0$ or $k+q \to 0$. This
 rules out a constant term in $\zeta$. The lowest-order term is then a product
of two momenta, so we must have $\zeta \propto  k (k+q)$. Therefore, the generic form of the magnon-holon interaction in the long-wavelength limit is indeed 
\begin{equation}
  \label{eq:1}
  H_{I}  = \alpha  \sum_{k, p, q} k(k+q) c^\dagger_{ p-q}
c^{\phantom\dagger}_{ p} B_{k+q}^\dagger B^{\phantom\dagger}_k.
\end{equation}
This result is summarized
in Fig.~\ref{couplingdiagram}.

\begin{figure}
\includegraphics[width=.7\columnwidth]{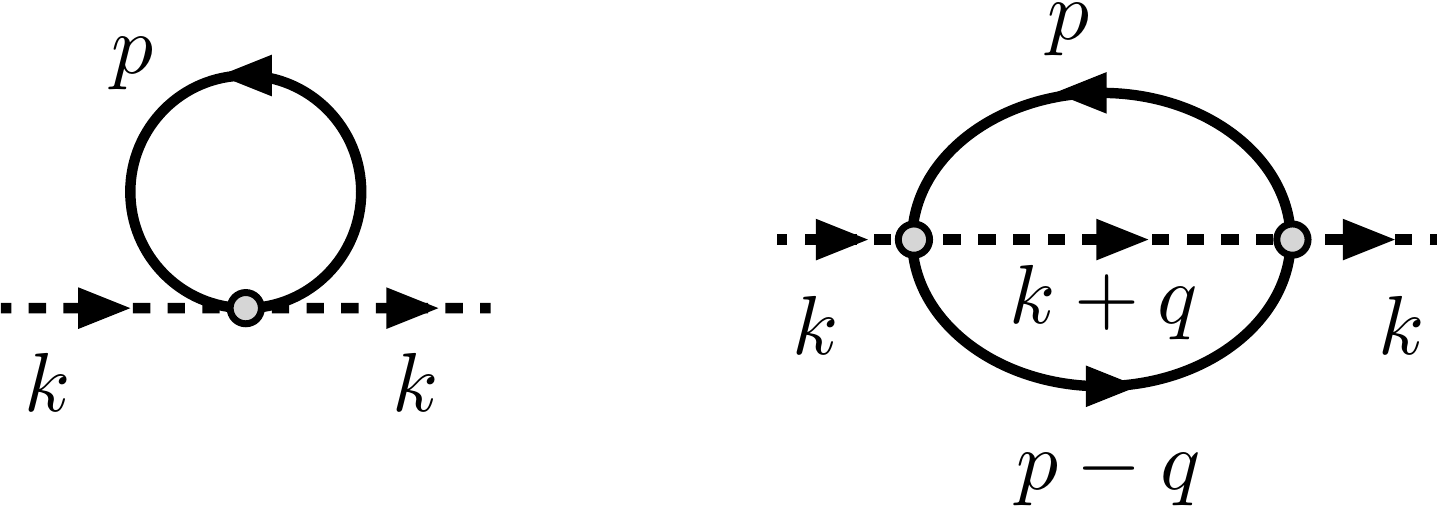}
\caption[selfenergy]{\label{selfenergy} First- and second-order
    diagrams in the magnon self-energy due to magnon-holon
  interaction.}
\end{figure}

Our complete Hamiltonian is $H=H_c+H_s+H_I$. To find out  how the interaction
affects the magnon spectrum, 
we calculate the magnon self-energy at first and  second order in $\alpha$. The corresponding diagrams are
illustrated in Fig.~\ref{selfenergy}. In the first-order (``tadpole'') diagram, the holon loop is 
proportional to the holon density and the interaction vertex
contributes with a factor of $k^2$. Therefore, the first order diagram only
renormalizes the magnon mass, with $\delta \lambda \sim \mathcal O( \alpha)$.

The second-order diagram is more interesting. To make the  
integrals well defined, we introduce a momentum cutoff $D$ around the
Fermi points $-p_F$ and $p_F$ for magnons and holons. We divide the self-energy into two contributions, $ \Sigma^{(2)} (k, \omega) =  \Sigma_0^{(2)} (k, \omega) + \Sigma_{2p_F}^{(2)}(k, \omega) $. Both contributions contain 
   logarithmic singularities in their real parts:
   \begin{eqnarray}
  \label{selfenergysecondorder}
  \Re \Sigma_0^{(2)}  &\sim& \alpha^2 \left( \omega -
    \lambda k^2\right) \log \left( \frac{D}{\omega - \lambda
      k^2} \right), \nonumber \\
\Re \Sigma_{2p_F}^{(2)} &\sim &\alpha^2  
[\omega - \lambda \left( k - 2 p_F \right)^2] \log \left[
      \frac{D}{\omega - \lambda \left(k - 2 p_F\right)^2}
    \right] \nonumber \\
&&+ (p_F\to -p_F).
\end{eqnarray}
The
 first contribution arises from low-momentum scattering and
is typical of the orthogonality catastrophe, which has already been
studied in the case of bosons.
\cite{Zvonarev:2007dx,Kamenev:2009ez} The second contribution
    corresponds to a momentum
transfer of $2 p_F$; it diverges on-shell only if $| k \pm 2 p_F| \approx 
|k|$, where $k$ is the magnon momentum. These singularities tell us some important mechanism is taking place
when magnons and holons scatter around the Fermi points.

\section{Effective field theory for magnon-holon interaction}
\label{sec:effect-field-theory}

In this section, we derive an effective field theory  for the scattering between  holons and magnons with momentum  close to the Fermi points $\pm p_F$.  Our goal is to enlighten the mechanism behind the infrared  singularities encountered in Sec.~\ref{sec:strongly-inter-regim} and to set the stage for going beyond perturbation theory using the numerical renormalization group in Sec.~\ref{sec:1}.

\subsection{Chirality Kondo effect\label{sec:chiralKondo}}
We start by rewriting the free Hamiltonians of Eq.~\eqref{hdemagnon} and
\eqref{Hcharge} in the continuum limit in terms of holon and magnon fields:
\begin{eqnarray}
  \label{Hcfield}
  H_c = \gamma \int dx~ \partial_x \psi^\dagger (x) \partial_x
  \psi (x), \\
  \label{Hsfield}
  H_s = \lambda \int dx~ \partial_x B^\dagger (x) \partial_x
  B (x).
\end{eqnarray}
where $\psi(x)=\frac{1}{\sqrt L}\sum_{p}c_pe^{ipx}$ and $B(x)=\frac{1}{\sqrt L}\sum_{k}B_ke^{ikx}$, with $L$ the system size.

We now restrict to low energies compared with the holon Fermi energy $ \lambda p_F^2$.  In this regime holons can only be  scattered  in the vicinity of the Fermi points. We   expand the holon fields in terms of
left movers and right movers,
\begin{equation}
 \psi(x) \approx e^{-i p_F x} \psi_L
(x) + e^{i p_F x} \psi_R (x).
\end{equation}
We also focus on magnon states with momentum near $\pm p_F$, and expand 
\begin{equation}
 B (x) \approx e^{-i p_F x} B_L
(x) + e^{i p_F x} B_R (x).
\end{equation}

We define two-component spinor fields which combine right and left movers\begin{equation}
  \Psi (x) = \left(
\begin{array}{c}
  \psi_R (x) \\ \psi_L (x)
\end{array}
\right) , \quad  \Omega(x) = \left(
\begin{array}{c}
B_R (x) \\ B_L (x)
\end{array} \right).\end{equation}
Linearizing both holon and magnon dispersion for $k,p\approx \pm p_F$ (and measuring the holon dispersion from the Fermi energy), we can write 
\begin{eqnarray}
  \label{Hcemspinor}
  H_c &\approx & \int dx~\Psi^\dagger(x) \left(- i v \tau^z \partial_x
\right) \Psi (x), \\
  \label{Hsemspinor}
  H_s &\approx & \int dx~\Omega^\dagger(x) \left[ \omega (p_F) - i u \tau^z \partial_x
\right] \Omega (x),
\end{eqnarray}
where $v = 2 \gamma p_F$ and $u = 2 \lambda p_F$ are  the holon and magnon
velocities, respectively,  and $\tau^z$ is the $z$ Pauli matrix in the internal  chirality space. As discussed in Sec.~\ref{sec:strongly-inter-regim}, we expect $\lambda\ll \gamma$ in the limit $J\ll s,t$. Thus, we are in a ``slow magnon'' regime    $u\ll v$.  The spinor representation suggests regarding 
  the chirality indices as pseudospins,  $R =~\uparrow$ and $L
=~\downarrow$ (not to be confused with the original electron spin $\sigma$).  We will see soon that this
picture is useful to interpret the interaction, but it is important to
keep in mind that the physical meaning of $R$ ($L$) is that the
particle carries momentum close to $p_F$ ($-p_F$).

\begin{figure}
\includegraphics[width=1\columnwidth]{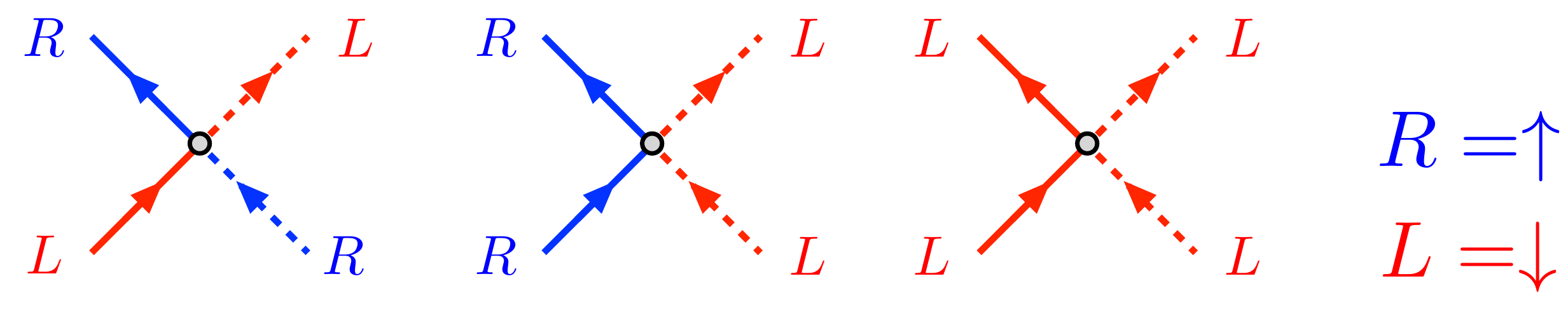}
\caption[pseudospin]{\label{pseudospin}  (color online) Different kinds of
  fermion-magnon scattering around the Fermi levels. The indices $L$
  and $R$ indicate whether the particle momentum is closest to $-p_F$ or
  $p_F$. By defining a chirality pseudospin such that $R =~\uparrow$
  and $L = ~\downarrow$, the scattering can be identified with
  pseudospin exchange interaction (as represented by the diagrams on
  the bottom). The solid line represent holons
  and the dashed line represent magnons. There are in fact six
  diagrams, but the remaining ones can be obtained from the ones
  presented through a parity transformation.}
\end{figure}

Now we must propose the form of the magnon-holon interaction. As
before, we appeal to symmetries. As we discussed in Sec.~\ref{sec:strongly-inter-regim}, conservation of magnetization implies that a local interacting term
must have the general form $\psi_{r_1}^\dagger(x) \psi^{\phantom\dagger}_{r_2} (x)
B^\dagger_{r_3} (x)B^{\phantom\dagger}_{r_4}(x) $  in order to conserve the number of
magnons. In this notation, $r$ represents a pseudospin index ($R$
or $L$). Momentum conservation impose constraints on the $r$'s:
only combinations such that $r_1 - r_2 + r_3 -
r_4 = 0$ are allowed. These are: $LLLL$ and
$RRRR$; $LLRR$ and $RRLL$; and $LRRL$ and $RLLR$. We split them into
pairs related by a parity transformation (where $L
\leftrightarrow R$). To ensure parity symmetry, coupling constants for
each element of a pair must be the same. The corresponding scattering
processes are illustrated in Fig.~\ref{pseudospin}.

With these restrictions, it is not difficult to show that, in the
spinor representation, the interacting part of the Hamiltonian assumes
the form
\begin{eqnarray}
  \label{HIrewritten}
H_I &=& V \int dx~ \Psi^\dagger   \Psi\Omega^\dagger  \Omega
 +\frac{ J_\parallel}{4} \int dx~ \Psi^\dagger \tau^z \Psi 
\Omega^\dagger \tau^z \Omega  \nonumber\\
&&+\frac{J_\perp}{8} \int dx~ ( \Psi^\dagger   \tau^+ \Psi  \,
  \Omega^\dagger   \tau^- \Omega  +\hc),
\end{eqnarray}
 where $V,J_\parallel,J_\perp$ are
three independent coupling constants. Since this interaction term
must arise from linearizing Eq.~\eqref{eq:1},   we expect the
coupling constants to be of the order of $\alpha p_F^2$.

In terms of pseudospins,
the first term in  Eq.~\eqref{HIrewritten} describes the
potential scattering between holons and magnons, while the remaining
terms describe an anisotropic exchange interaction between holon
and magnon pseudospins. Recognizing $\mathbf{S}_F (x) \equiv  \Psi^\dagger(x) \frac{\pmb{\tau}}{2}\Psi(x)$ and
$\mathbf{S}_M (x) \equiv  \Omega^\dagger(x) \frac{\pmb{\tau}}{2} \Omega (x)$ as pseudospin vector densities, we
further rewrite $H_I$ as
\begin{eqnarray}
  \label{HIrewrittenspin}
  H_I &=& V \int dx~ \Psi^\dagger  \Psi  \Omega^\dagger 
  \Omega  
+J_\parallel \int dx~  S_{F}^z (x)  S_{M}^z (x) \nonumber\\
&&+\frac{J_\perp}{2} \int dx~ \left[ S_{F}^+ (x)  S_{M}^- (x) +
\hc\right], 
\end{eqnarray}
where the label $F$ is for (fermionic) holons and the label $M$ is for magnons.

The Hamiltonian in Eq. \eqref{HIrewrittenspin} can be viewed as a Kondo interaction\cite{Hewson:1997vc} between a finite density of fermions (represented by the holons) and a   number of \emph{mobile} impurities (represented by the magnons). The connection with the Kondo model is made more clear in the limit of vanishing magnon density (which is the interesting regime to investigate the stability of the ferromagnetic state against infinitesimal perturbations). Since magnetization is
conserved, we may restrict our problem to the subspace with  a single magnon. In this case we can rewrite the Hamiltonian using first quantization for magnon operators:\begin{eqnarray}
H_s& =&\omega(p_F)+2uS_M^z \hat P,\label{Hsemspinor1stq}\\
H_I&=&V  \Psi^\dagger(\hat X)  \Psi  (\hat X)+J_\parallel    S_{F}^z (\hat X)  S_{M}^z  \nonumber\\
&&+\frac{J_\perp}{2}   \left[ S_{F}^+ (\hat X)  S_{M}^- +
\hc\right], 
\end{eqnarray}
where $\hat X$ and $\hat P$ are the position and momentum operators of the magnon, respectively, obeying $[\hat X,\hat P]=i$. Note that the Hamiltonian is translationally invariant, but the interaction is local at the position of the magnon.  This kind of dynamics has  been 
encountered in the coupling of a mobile impurity with a
Luttinger liquid.\cite{Lamacraft:2009bp}

Next, we apply a Galilean  transformation to the magnon reference frame.\cite{Pines:PhysRev.90.297,Neto:1996tr} 
The transformation  is
\begin{equation}
  \label{transfcanonica}
  U \equiv e^{-i \hat X\int dx\, \Psi^\dagger (-i\partial_x)\Psi},
\end{equation}
and is such that  \begin{eqnarray}
\tilde \Psi(x)&=&U^\dagger \Psi(x)U= \Psi(x-\hat X),\\
\tilde P&=&U^\dagger \hat PU=\hat P -\int dx\, \Psi^\dagger (-i\partial_x)\Psi.
\end{eqnarray}
As a result, the transformed interaction $\tilde{H}_I = U^\dagger H_I U$ becomes
\begin{eqnarray}
  \label{HItransformed}
 \tilde{H}_I &=& V  \Psi^\dagger (0) \Psi(0) +J_\parallel   S^z_{F} (0)  S^z_{M}  \nonumber\\
&&+\frac{J_\perp}{2}  \left[ S^+_{F} (0)  S^-_{M}  +
\hc\right]. 
\end{eqnarray}
The magnon-holon interaction is now restricted to $x = 0$, since we are in a frame whose origin moves along with the magnon. For consistency,  we must also transform the
kinetic energy terms $H_c$ and $H_s$. The free holon Hamiltonian remains unchanged, i.~e., $\tilde{H}_c = H_c$. However,  the magnon term in   Eq.~(\ref{Hsemspinor1stq}) gets modified:
\begin{eqnarray}
  \label{Hsmodified}
    \tilde{H}_s &=& U^\dagger H_s U  \nonumber\\
    &=& \omega(p_F)+2u S_M^z\hat P+2iuS_M^z\int dx\, \Psi^\dagger \partial_x\Psi.
\end{eqnarray}

At this point we have  eliminated the magnon position $\hat X$ from the transformed Hamiltonian $\tilde H=\tilde H_c+\tilde H_s+\tilde H_I$. Therefore, $[ \tilde{H}, \hat P ] = 0$, implying that $\tilde{H} $
and $\hat P$ can be simultaneously diagonalized. This is a consequence of translational invariance of the original Hamiltonian.\cite{Pines:PhysRev.90.297,Neto:1996tr} We shall focus on the subspace with       eigenvalue    $P = 0$, which corresponds to  a magnon with energy exactly equal to $\omega(p_F)$. In this case the first term in  Eq. (\ref{Hsmodified}) vanishes. Deviations from
the $P=0$ condition  amount to an extra term proportional to $S_M^z$, which is equivalent to  an effective magnetic field acting on the magnon pseudospin. While we have got rid of $\hat X$, the price we paid is the introduction of the second  term in Eq. (\ref{Hsmodified}). This term couples the magnon pseudospin to the total momentum operator for holons.

We can now add $\tilde{H}_s$ for $P=0$ to the free holon Hamiltonian $\tilde H_c$:
\begin{eqnarray}
  \tilde{H}_c + \tilde{H}_s &=& \omega(p_F)+\int dx\, \psi^\dagger_R(v-2uS_M^z)(-i\partial_x)\psi^{\phantom\dagger}_R\nonumber\\
  && +\int dx\, \psi^\dagger_L(v+2uS_M^z)(i\partial_x)\psi^{\phantom\dagger}_L.
  \label{freeHaftertransformation} 
  \end{eqnarray}
We further
rewrite the right-hand side of Eq.~\eqref{freeHaftertransformation} by taking $\psi_L(x)\to \psi_L(-x)$ (this is equivalent to relabeling holon states with momentum $p$ as $-p$
for the left branch):
\begin{equation}
  \label{finaldispersionfermion}
\tilde{H}_c + \tilde{H}_s = \omega(p_F)+ \int dx\, \Psi^\dagger  ( v -  2uS^z_{M}\tau^z )(-i\partial_x)\Psi.
\end{equation}
We emphasize that $\tau^z$ in Eq. (\ref{finaldispersionfermion}) acts in the holon pseudospin space. 
Recall that $v$ is the holon velocity around the Fermi points $\pm p_F$. Eq.~\eqref{finaldispersionfermion} shows that the holon 
velocity in the transformed Hamiltonian is now given by  $v\pm u$ depending on whether   the holon and magnon
have the same or opposite pseudospins, i.~e. move  in the same or opposite directions.

Finally,  dropping the constant $\omega(p_F)$ and adding  the transformed  interaction of
Eq.~\eqref{HItransformed}, we arrive at our effective Kondo-type Hamiltonian for the scattering of a single magnon with momentum close to $\pm p_F$ by low-energy holons: 
\begin{eqnarray} 
  \tilde H &=&  \sum_{k=-\infty}^{+\infty}  \Psi_k^\dagger \left( v  - 2 u S^z_{M} \tau^z
  \right)k ~\Psi_k\nonumber\\
  &&+ \frac{J_{\perp}}{2} [ S^+_{F}(0)
    S^-_{M} + \hc]  \nonumber \\
&&+ J_{\parallel} S^z_{F} (0)    S^z_{M} 
+ V \Psi^\dagger (0) \Psi (0)\label{eq:1}.
\end{eqnarray}

\subsection{Perturbative renormalization group analysis\label{sec:pertRG}}

If the magnon had zero velocity, i.~e., $u = 0$, the model of
Eq.~\eqref{eq:1} would correspond  precisely to the  Kondo
model, which describes a Fermi sea of electrons coupled to a spin-$1/2$ 
impurity   at the origin through an anisotropic exchange interaction.
\cite{Hewson:1997vc} Here the model also includes the potential scattering $V$, which is generally present in the absence of particle-hole symmetry. It is well known that perturbation theory in the Kondo interaction gives rise to infrared divergences like the ones we encountered in Eq.~\eqref{selfenergysecondorder}. The Kondo model has been investigated non-perturbatively using the    numerical renormalization group (NRG) \cite{KWW80:1003,BCP:2007}
and solved exactly  through Bethe ansatz.\cite{Andrei:1983tl} From the renormalization group (RG) point of view, the logarithmic  singularities that arise in perturbation theory can be recast in the renormalization of effective coupling constants at the appropriate energy scale. For $u=0$,   the exchange coupling constants flow to strong coupling if $J_\perp> -J_\parallel$. The low-energy fixed point of the model can   be understood in terms of the formation of a  singlet between the localized  impurity and one electron from the conduction band, which  decouples from the
remaining electrons. In other words, the fermion spin screens the impurity spin.

In our case, the renormalization of the effective exchange interactions decides the fate of the magnon in the low-energy limit. We first approach the problem by applying a perturbative RG approach to interaction  \eqref{HIrewritten} with $u>0$. The procedure is similar to Anderson's poor man's scaling,\cite{Anderson:0022-3719-3-12-008} generalized to include a momentum cutoff in the magnon sub-band and to respect total momentum conservation in the virtual magnon-holon scattering process. We obtain the set of RG equations: \begin{eqnarray}
\frac{dJ_\perp}{d\ell}&=&\frac{vJ_\perp (J_\parallel-V)}{2\pi(v^2-u^2)},\label{RG1}\\
\frac{d(J_\parallel-V)}{d\ell}&=&\frac{J_\perp^2}{2\pi(v+u)},\label{RG2}\\
\frac{d(J_\parallel+V)}{d\ell}&=&\frac{J_\perp^2}{2\pi (v-u)},\\
\frac{du}{d\ell}&=&-\frac{uJ_\perp^2}{8\pi (v^2-u^2)},\label{RG4}
 \end{eqnarray}
where $d\ell =d\Lambda/\Lambda$ denotes the infinitesimal reduction of the ultraviolet cutoff in the RG step. 

To analyze the RG flow, we first note that to second order in perturbation theory the renormalization of $J_\parallel +V$ is controlled by $J_\perp$ and does not feed back into the beta functions for  $J_\perp$ and $J_\parallel-V$. We can then focus on Eqs. \eqref{RG1} and \eqref{RG2}. In the slow magnon regime $u<v$, these coupled equations define a Kosterlitz-Thouless type flow diagram in the $(J_\perp,J_\parallel-V)$ plane, with a  flow to strong coupling for $J_\perp>-(J_\parallel-V)$. Furthermore, Eq. \eqref{RG4} implies that the magnon velocity decreases with the RG flow. Here we should remark  that we can follow the flow to arbitrarily low energies only in the case $P=0$, i.~e. when the magnon momentum exactly matches the holon Fermi momentum $p_F$.  As discussed in   Sec.~\ref{sec:chiralKondo}, deviations from this condition are equivalent to an effective magnetic field coupled to the magnon pseudospin, which   cuts off the RG flow at the  scale set by   $P$.

The enhancement of the effective couplings   suggests that in the low-energy limit the magnon is strongly backscattered between   states with momentum $\pm p_F$, and its group velocity may eventually vanish as suggested by Eq. \eqref{RG4}. By analogy with the original Kondo effect, we expect that at the low-energy fixed point the magnon will form a ``chirality  singlet''  with one holon, as represented  by the state  $|\Phi\rangle =\frac1{\sqrt2}\int dx[B^\dagger_R(x)\psi^\dagger_L(x)-B^\dagger_L(x)\psi^\dagger_R(x)]|0\rangle$. This singlet involves a pair of (distinguishable) particles which are delocalized in space and move with opposite momenta ---  somewhat akin to a Cooper pair in BCS theory. Note that the particle-hole symmetry of our model is broken both by  $V\neq 0$ and $u\neq 0$; thus, the putative low-energy fixed point should contain a marginal potential scattering operator that accounts for  a non-universal phase shift in the remaining holon states. In   Sec. \ref{sec:1} we shall investigate the nature of the low-energy fixed point using the NRG method.

\section{Numerical renormalization group}
\label{sec:1}

The numerical diagonalization of the Hamiltonian~(\ref{eq:1}) follows the standard NRG method. The
procedure having been thoroughly detailed, \cite{KWW80:1003,BCP:2007} only cursory description is
necessary, meant to prepare the discussion of the flow in renormalization-group space that
constitutes the central object of this section.

\subsection{Procedure}
\label{sec:2}
To diagonalize the Hamiltonian~\eqref{eq:1}, we will rely on the numerical renormalization-group
procedure, an approximate iterative method that depends on strictly controllable approximations to yield a set
of eigenvalues and eigenvectors from which essentially exact physical properties. More important
from our perspective, the iterative diagonalization can be regarded as a sequence of
renormalization-group transformations and therefore accurately describes the flow of the Hamiltonian
in renormalization-group space.

To define the renormalization-group transformation, we need a scaled truncated version of the
Hamiltonian. Our initial goal is to project Eq.~(\ref{eq:1}) on a finite basis. Relative to the
basis of the holon and magnon states, the new basis will of course be incomplete. Special care
will be taken to preserve the magnon states and their interaction with the holons.  Only the
first term on the right-hand side of Eq.~(\ref{eq:1}) will be affected by the projection, which
comprises three steps.

\subsubsection{Logarithmic discretization of the conduction band}
\label{sec:3}
The first step is controlled by an arbitrary dimensionless parameter $\Lambda>1$. Given $\Lambda$,
we split the conduction band into two segments, one above and the other below the Fermi level,
 and divide each segment into an infinite logarithmic sequence of intervals
 \begin{align}
   \mathcal{I}_{m\pm}=\pm D\big(\Lambda^{-m-1}, \Lambda^{-m}\big)\qquad(m=0,1,\ldots),
 \end{align}
where the positive segment runs from $\epsilon_{k}=\epsilon_{F}\equiv0$
to $\epsilon_{k}=D$, and the negative one from $\epsilon_{k}=-D$ to $\epsilon_{k}=0$.

For each interval $\mathcal{I}_{m\pm}$, we define the normalized Fermi operator
\begin{align}
  a_{m\pm} = \dfrac{\Lambda^{m/2}}{\sqrt{1-\Lambda^{-1}}}
\sum_{k\in \mathcal{I}_{m\pm}}\Psi_{k}\qquad(m=0,1,\ldots).
  \label{eq:2}
\end{align}

It follows that the operator $\Psi(0)$ is a linear combination of the $a_{m\pm}$:
\begin{align}
\Psi(0) = 
\sqrt{1-\Lambda^{-1}}\sum_{m=0}^{\infty}\Lambda^{-m/2}\left(a_{m+}+a_{m-}\right)\equiv\sqrt2\f{0},
\label{eq:3}
\end{align}
where we have defined the normalized Fermi operator $\f{0}$, so that Eq.~(\ref{eq:1}) can be written
in the exact form
\begin{align}
  \tilde H = &\sum_{k}\Psi_{k}^{\dagger}(v -2uS^{z}_{M}\tau^{z})\Psi_{k}+
J_{\perp}\left(\fd{0\uparrow}\f{0\downarrow}S^-_{M}+\hc\right) \nr
&+ 2J_{\parallel}\left(\fd{0\uparrow}\f{0\uparrow}-\fd{0\downarrow}\f{0\downarrow}\right)S^z_{M}\nonumber\\
&+2\vv\left(\fd{0\uparrow}\f{0\uparrow}+\fd{0\downarrow}\f{0\downarrow}\right).&
\label{eq:4}
\end{align}
Here $\uparrow=R$ and $\downarrow =L$ refer to the chirality pseudospins defined in Sec. \ref{sec:effect-field-theory}. From now on we shall call these simply ``holon spins'', whereas $\mathbf S_M$ is  the ``magnon spin''. 

Next, we project the first term on the right-hand side of Eq.~(\ref{eq:4}) on the basis of the
$\ampm$, which yields the approximate equality
\begin{eqnarray}
  \tilde H &= &\sum_{m}\mathcal{E}_{m}
  \left[v(\amd{m\uparrow}\am{m\uparrow}+\amd{m\downarrow}\am{m\downarrow})
\right.\nonumber\\&
& \left.-2u(\amd{m\uparrow}\am{m\uparrow}-\amd{m\downarrow}\am{m\downarrow})S^z_{M}\right] \nonumber\\
&&+ 
J_{\perp}\left(\fd{0\uparrow}\f{0\downarrow}S^-_{M}+\hc\right)\nonumber\\
&&+ 2J_{\parallel}\left(\fd{0\uparrow}\f{0\uparrow}-\fd{0\downarrow}\f{0\downarrow}\right)S^z_{M}\nonumber\\
&&
+2\vv\left(\fd{0\uparrow}\f{0\uparrow}+\fd{0\downarrow}\f{0\downarrow}\right),
\label{eq:5}\end{eqnarray}
where
\begin{align}
\mathcal{E}_{m}= \dfrac{1-\Lambda^{-1}}{\log\Lambda}\Lambda^{-m}
\label{eq:6} 
\end{align}
is the average momentum in the interval $\mathcal{I}_{m}$.\cite{CO05.104432} 

The projection on the incomplete basis introduces deviations in computed physical properties of
$\mathcal{O}[\exp(-\dfrac{\pi^{2}}{\log\Lambda})]$,\cite{KWW80:1003} which are insignificant for
$\Lambda \alt 3$.

\subsubsection{Lanczos transformation}
\label{sec:5}
Numerical treatment of the Hamiltonian~\eqref{eq:5} calls for truncation of the infinite basis. To
preserve the interaction terms on the right-hand side, before truncation we define a new infinite
basis $\{\f{n}\}$ ($n=0,1,\ldots$), where the Fermi operators $\f{n}$ form an orthonormal sequence of linear
combinations of the $\am{m\pm}$,
\begin{align}
  \label{eq:7}
  \f{n} = \sum_{m=0}^{\infty}\alpha_{nm}\Big[a_{m+}+(-1)^{n}a_{m-}\Big].
\end{align}

In the new basis, $\f{0}$ is the operator defined by Eq.~\eqref{eq:3},
so that
\begin{align}
  \label{eq:8}
  \alpha_{0m} = \sqrt{\dfrac{1-\Lambda^{-1}}2}\Lambda^{-m/2},
\end{align}
and the remaining Fermi operators are tailored to the requirement that
\begin{align}
  \sum_{m=0}^{\infty} \mathcal{E}_{m}(\amd{m+}\am{m+}-\amd{m-}\am{m-})
  = \sum_{n=0}^{\infty}t_{n}(\fd{n}\f{n+1}+\hc).
 \label{eq:9}
\end{align}

Wilson has shown that Eq.~\eqref{eq:9} is satisfied with\cite{Wi75:773}
\begin{align}
  \label{eq:10}
  t_{n} =
 \dfrac{ \mathcal{E}_{0}(1-\Lambda^{-n-1})\Lambda^{-n/2}}{\sqrt{(1-\Lambda^{-2n-1})(1-\Lambda^{-2n-3})}}
  \qquad(n=0,1,\ldots)
\end{align}
and determined the coefficients $\alpha_{nm}$. In particular, for $n=1$,
\begin{align}
  \label{eq:11}
  \alpha_{1m} = \sqrt{\dfrac{1-\Lambda^{-3}}2}\Lambda^{-3m/2}.
\end{align}

We now substitute the \rhs\ of Eq.~\eqref{eq:9} for the kinetic terms on the \rhs\ of
Eq.~\eqref{eq:5}, to obtain the Lanczos transformed Hamiltonian
\begin{align}
 \tilde H =& \sum_{n=0}^{\infty}t_{n} \left[
v(\fd{n\uparrow}\f{n+1\uparrow}+\fd{n\downarrow}\f{n+1\downarrow})\right.\nr &
\left.-2u(\fd{n\uparrow}\f{n+1\uparrow}-\fd{n\downarrow}\f{n+1\downarrow})S^z_{M}+\hc \right]&
\nr
&+ 
J_{\perp}\left(\fd{0\uparrow}\f{0\downarrow}S^-_{M}+\hc\right)&\nr
&+ 2J_{\parallel}\left(\fd{0\uparrow}\f{0\uparrow}-\fd{0\downarrow}\f{0\downarrow}\right)S^z_{M}&\nr
&+2\vv\left(\fd{0\uparrow}\f{0\uparrow}+\fd{0\downarrow}\f{0\downarrow}\right),
  \label{eq:12}
\end{align}
which leaves us in position to truncate the infinite basis.

\subsubsection{Truncation and scaling}
\label{sec:6}
Inspection of Eq.~\eqref{eq:10} shows that the fraction on the \rhs\ rapidly approaches unity as $n$
grows, so that to an excellent approximation,
\begin{align}
  \label{eq:13}
  t_{n} = \mathcal{E}_{0}\Lambda^{-n/2}\qquad(n\ \mathrm{such\ that}\ \Lambda^{-n}\ll 1).
\end{align}

If we are interested in physical properties at the energy scale $E$, it is safe to truncate the infinite
series on the \rhs\ of Eq.~\eqref{eq:12} at $n=N$, where $N$ is the smallest integer satisfying
$t_{N} <\epsilon_{IR} E$ for a specified dimensionless \emph{infrared cutoff} $\epsilon_{IR}^{\phantom{\dagger}}\ll
1$. Given an energy $E$ and an infrared cutoff $\epsilon_{IR}^{\phantom{\dagger}}$, we define the truncated, scaled Hamiltonian
\begin{align}
  \label{eq:14}
  H_{N} \equiv\dfrac1{\drg{N-1}}
\Bigg\{&\sum_{n=0}^{N-1}t_{n}\left[
v(\fd{n\uparrow}\f{n+1\uparrow}+\fd{n\downarrow}\f{n+1\downarrow})
\right.\nr 
&\left.-2u(\fd{n\uparrow}\f{n+1\uparrow}-\fd{n\downarrow}\f{n+1\downarrow})S^z_{M}+\hc\right]
\nr
&+J_{\perp}\left(\fd{0\uparrow}\f{0\downarrow}S^-_{ M}+\hc\right)&\nr
&+ 2J_{\parallel}\left(\fd{0\uparrow}\f{0\uparrow}-\fd{0\downarrow}\f{0\downarrow}\right)S^z_{M}&\nr
&+2\vv\left(\fd{0\uparrow}\f{0\uparrow}+\fd{0\downarrow}\f{0\downarrow}\right)
\Bigg\},
\end{align}
where $\drg{n}$ denotes the right-hand side of Eq.~\eqref{eq:13}, i.~e., $
  \drg{n} \equiv \mathcal{E}_{0}\Lambda^{-n/2}$.

\subsubsection{Iterative diagonalization}
\label{sec:7}

The truncated form~\eqref{eq:14} is convenient for iterative diagonalization.\cite{KWW80:1003} With
$N=1$, only the magnon spin and the operators $\f{0}$ and $\f{1}$ contributing to the right-hand side, the 
Hamiltonian $H_{N=1}$ is equivalent to a $32\times32$ matrix that is easily diagonalized,
numerically. The next Hamiltonian in the iterative procedure, $H_{N=2}$ is then projected on the
basis $\ket{r}_{1}$, $\fd{N+2\uparrow}\ket{r}_{1}$, $\fd{N+2\downarrow}\ket{r}_{1}$,
$\fd{N+2\uparrow}\fd{N+2\downarrow}\ket{r}_{1}$ ($r=1,2,\ldots,32$), where $\ket{r}_{1}$ is one of the
eigenstates resulting from the diagonalization of $H_{N=1}$, and the resulting $128\times128$ matrix
is numerically diagonalized. This completes the first iterative cycle ($N=2$).

More generally, given the eigenstates $\ket{r}_{N}$ of $H_{N}$, the Hamiltonian $H_{N+1}$ is
projected on the basis resulting from the operators $\mathds{1}$, $\fd{N+1\uparrow}$,
$\fd{N+1\downarrow}$, and $\fd{N+1\uparrow}\fd{N+1\downarrow}$ applied to the $\ket{r}_{N}$ and
diagonalized. The number of eigenstates of $H_{N}$ in the first few iterations is
$N_{r}=2^{1+2N}$. To keep the rapidly growing $N_{r}$ under control, only the eigenstates of $H_{N}$
corresponding to eigenvalues below a fixed \emph{ultraviolet cutoff} $E_{uv}$ are computed at
each iteration, so that the computational cost grows linearly, instead of exponentially, with the
number of iterations. The ultraviolet cutoff $E_{uv}$, the discretization parameter $\Lambda$ and
the infrared cutoff $\epsilon_{IR}^{\phantom{\dagger}}$ control the computational cost and the
accuracy of the physical properties calculated from the eigenstates and eigenvalues of $H_{N}$.

\subsubsection{Renormalization-group transformation}
\label{sec:8}

The factor $1/\drg{N-1}$ on the right-hand side of Eq.~\eqref{eq:14} expresses the Hamiltonian $H_{N}$
in units of its smallest matrix element, $t_{N-1}\approx \drg{N-1}$ and defines a
renormalization-group transformation $\mathcal{T}$, which turns a scaled truncated Hamiltonian into
another scaled Hamiltonian, one that is truncated at a finer scale. More specifically, from
Eq.~\eqref{eq:14} we have that 
\begin{align}
  H_{N+2} =& \Lambda H_{N} +
  \sum_{n=N}^{N+1}\dfrac{t_{n}}{\drg{N+1}}\Big[v(\fd{n\uparrow}\f{n+1\uparrow}+\fd{n\downarrow}\f{n+1\downarrow})&\nr
    &-2u(\fd{n\uparrow}\f{n+1\uparrow}-\fd{n\downarrow}\f{n+1\downarrow})S^z_{M}+\hc\Big],  \label{eq:16}&
\end{align}
which defines the transformation $\trg{H_{N}} = H_{N+2}$.

At first sight, it may seem more natural to step $N\to N+1$. Nonetheless, the $H_{N}\to H_{N+1}$
transformation is unwieldy, because successive applications almost invariably generate two-point
limit cycles.\cite{KWW80:1003} The $H_{N}\to H_{N+2}$ transformation, by contrast, has simple fixed
points.

\subsection{Fixed points}
\label{sec:9}
As first shown by Wilson,\cite{Wi75:773} the spin-flip term on the \rhs\ of Eq.~\eqref{eq:1} is a
marginally relevant operator. To identify the fixed points of $\trg{H_{N}}$ we will therefore
consider two forms of the Hamiltonian that remain invariant under scaling of $J_{\perp}$. For weak
coupling between the magnon and the holons, for small $N$ the scaled truncated
Hamiltonian $H_{N}$ in Eq.~\eqref{eq:14} tends to be close to one of them. As $N$ grows, our
numerical analysis shows that $H_{N}$ flows towards the second fixed point. We   will therefore
refer to the former (latter) as the \emph{weak-coupling} (\emph{strong-coupling}) fixed points. 

\subsubsection{Weak-coupling fixed points ($J_{\perp}=0$)}
\label{sec:4}
Without the spin-flip term, the Hamiltonian~\eqref{eq:14} reads
\begin{align}
  \drg{N-1}H_{N}^{0} =&
 \sum_{n=0}^{N-1}t_{n}\left[
v(\fd{n\uparrow}\f{n+1\uparrow}+\fd{n\downarrow}\f{n+1\downarrow})\right.&\nr
&\left.-2u(\fd{n\uparrow}\f{n+1\uparrow}-\fd{n\downarrow}\f{n+1\downarrow})S^z_{M}+\hc\right]&
\nr
& 
+ 2J_{\parallel}\left(\fd{0\uparrow}\f{0\uparrow}-\fd{0\downarrow}\f{0\downarrow}\right)S^z_{M}
&\nr
&+2\vv\left(\fd{0\uparrow}\f{0\uparrow}+\fd{0\downarrow}\f{0\downarrow}\right)
 ,
  \label{eq:17}
\end{align}
a form that commutes with $S^z_{M}$.

The spectrum of $H_{N}^{0}$ can therefore be classified by the $z$ component $s^{z}_{M}$ of the
magnon spin. It can be divided, in other words, into an $s^{z}_{M}=1/2$ sector and another with
$s^{z}_{M}=-1/2$.

For each magnon-spin component, the Hamiltonian in Eq.~\eqref{eq:17} splits into an $\uparrow$
and a $\downarrow$ terms, labeled by the holon spin. For $s^{z}_{M}= +1/2$, we have that
\begin{align}
  \label{eq:18}
  H_{N}^{0} = H_{N,\uparrow}^{0+} + H_{N,\downarrow}^{0+},
\end{align}
where
  \begin{align}
    \label{eq:19}
    \drg{N-1}H_{N,\uparrow}^{0+} = &
(v-u)\sum_{n=0}^{N-1}t_{n}(\fd{n\uparrow}\f{n+1\uparrow}+\hc)&
\nr& + (2\vv+J_{\parallel})\fd{0\uparrow}\f{0\uparrow}&
  \end{align}
and
\begin{align}
  \label{eq:20}
    \drg{N-1}H_{N,\downarrow}^{0+} =& 
(v+u)\sum_{n=0}^{N-1}t_{n}(\fd{n\downarrow}\f{n+1\downarrow}+\hc)&\nr 
&+ (2\vv-J_{\parallel})\fd{0\downarrow}\f{0\downarrow}.&
\end{align}

With $s^{z}_{M}=-1/2$, the Hamiltonian~\eqref{eq:18} likewise splits in two terms:
\begin{align}
  \label{eq:21}
  H_{N}^{0} = H_{N,\uparrow}^{0-} + H_{N,\downarrow}^{0-},
\end{align}
where
\begin{align}
  \label{eq:22}
    \drg{N-1}H_{N,\uparrow}^{0-} = &
(v+u)\sum_{n=0}^{N-1}t_{n}(\fd{n\uparrow}\f{n+1\uparrow}+\hc)&\nr
&+ (2\vv-J_{\parallel})\fd{0\uparrow}\f{0\uparrow},&
\end{align}
and
\begin{align}
  \label{eq:23}
    \drg{N-1}H_{N,\downarrow}^{0-} = &
(v-u)\sum_{n=0}^{N-1}t_{n}(\fd{n\downarrow}\f{n+1\downarrow}+\hc)&\nr
&+ (2\vv+J_{\parallel})\fd{0\downarrow}\f{0\downarrow}.&
\end{align}
Comparison of the \rhs s of Eqs.~\eqref{eq:19}~and \eqref{eq:23} shows that $H_{N,\uparrow}^{0+}$ and
$H_{N,\downarrow}^{0-}$ transform into each other under the inversion
$\f{n\uparrow}\leftrightarrow\f{n\downarrow}$ and therefore have the same spectrum. Likewise, $H_{N,\uparrow}^{0-}$ and
$H_{N,\downarrow}^{0+}$ transform into each other under the same inversion and have identical spectra. 

In fact, the four Hamiltonians, $H_{N,\uparrow}^{0\pm}$ and $H_{N,\downarrow}^{0\pm}$, have the same
general form
\begin{align}
  \label{eq:24}
  H^{D,\ww}_{N}= D\sum_{n=0}^{N-1}t_{n}(\fd{n}\f{n+1}+\hc)+\ww\fd{0}\f{0},
\end{align}
where $D$ and $\ww$ are constants.

It has long been established that, as $N$ grows so that $N\to N+2$, the Hamiltonian $H_{D,\ww}^{N}$
rapidly approaches a single-particle fixed point labeled by a phase shift $\delta$.\cite{KWW80:1044}
Although the fixed-point Hamiltonians depends on the parity of $N$, the structures for
even and for odd $N$ are similar. If $N$ is even, for example, the fixed-point Hamiltonian
is given by the expression
\begin{align}
  \label{eq:25}
  H_{\delta}^{*}= D\left[\eta_{0}g_{0}^{\dagger}g_{0}
+\sum_{m=1}^{\infty}(\eta_{m+}g_{m+}^{\dagger}g_{m+}^{\phantom\dagger}-\eta_{m-}g_{m-}^{\dagger}g_{m-}^{\phantom\dagger})\right],
\end{align}
where $\eta_{0}$ is a number between $-\sqrt\Lambda$ and $\sqrt\Lambda$ that has to be determined
numerically, and to a good approximation (within $\mathcal{O}(\Lambda^{-2m})$ relative deviation),
\begin{align}
  \label{eq:26}
  \eta_{m\pm} = \Lambda^{m-\frac12\mp\frac{\delta}{\pi}}\qquad(m=1,2,\ldots)
\end{align}
with a phase shift $\delta$ dependent on the ratio $\ww/D$:
\begin{align}
  \label{eq:27}
  \tan\delta = -\pi \dfrac{\ww}{2D}.
\end{align}

Equations~\eqref{eq:25}, \eqref{eq:26},~and \eqref{eq:27} show that, as the even integer $N$ grows
to infinity, the scaled truncated Hamiltonian $H_{N}^{0}$ flows to a fixed point comprising two
infinite sets of eigenvalues $\{\varepsilon_{m\pm}^+\}$ and $\{\varepsilon_{m\pm}^-\}$
($m=0,1,2,\ldots$) associated with holon spins that are parallel or antiparallel to
the magnon spin, respectively.

The eigenvalues $\varepsilon_{m\pm}^{\pm}$ form the infinite logarithmic sequence
\begin{align}
  \label{eq:28}
  \varepsilon_{m\pm}^{\pm} = (v-u)\Lambda^{m-\frac12-\frac{\delta_{\pm}}{\pi}},
\end{align}
where
\begin{align}
  \label{eq:29}
  \tan\delta_{\pm}= -\pi\dfrac{\vv \pm J_{\|}/2}{v\mp u}
\end{align}

Since the phase shifts $\delta_{\pm}$ can take any value from $-\pi/2$ to $\pi/2$, depending on $V$
and $J_{\perp}$, Eq.~\eqref{eq:25} defines a line of (weak-coupling) fixed points.
Physically, the two sequences corresponds to the two bands discussed in Sec.~\ref{sec:effect-field-theory}, with
velocities  $v\pm u$ in the magnon reference frame. Each band is phase shifted. The phase shift is $\delta_{+}$ for holon spins that
are parallel to the magnon spin, and $\delta_{-}$ for spins that are antiparallel. Since
$\delta_{+}\ne\delta_{-}$ each flip of the magnon spin triggers two Anderson orthogonality catastrophes, one for
each component of the holon spin. We will come back to this issue in Sec.~\ref{sec:21}. Before
that, however, we have to examine the other set of fixed points.

\subsubsection{Strong-coupling fixed points ($J_{\perp}\to\infty$)}
\label{sec:10}
As the coupling $J_\perp$ on the \rhs\ of Eq.~\eqref{eq:14} grows, the magnon and the
$\f{0}$ orbital lock into a singlet, with energy $J_{\perp}$ below the other configurations of the
magnon $\f{0}$-orbital pair. As $J_{\perp}\to\infty$, the other configurations cannot contribute to
the physical properties of the Hamiltonian. The degrees of freedom associated with both the magnon
and $\f{0}$ are lost at this \emph{strong-coupling} fixed point. At iteration $N$ the basis is
therefore reduced to the operators $\f{n\mu}$ ($n=1,\ldots,N$, $\mu=\uparrow,\downarrow$). The
scaled truncated Hamiltonian can again be split into two decoupled components, i.~e., we have that
\begin{align}
  \label{eq:32}
  H_{N}^{\infty}=H_{N,\uparrow}^{\infty} + H_{N,\downarrow}^{\infty},
\end{align}
where both components are of the general form~\eqref{eq:24}. Specifically, we have that
\begin{align}
  \label{eq:33}
  H_{N,\mu}^{\infty}=
  D_{\infty}\sum_{n=0}^{N-2}t_{n}(\fd{n\mu}\f{n+1\mu}+\hc)+\ww_{\infty}\fd{0\mu}\f{0\mu},
\end{align}
where $\mu=\uparrow,\downarrow$ and $D_{\infty}$ and $\ww_{\infty}$ are constants that must be extracted from the iterative 
diagonalization of $H_{N}$.

The series on the \rhs\  of Eq.~\eqref{eq:33} runs from 0 to $N-2$ because we have shifted the
summation index $n\to n+1$. The coefficients $D_{\infty}$ and $\ww_{\infty}$ are independent of the
spin index $\mu$, because $H^{\infty,*}$ must have the symmetry of $H_{N}$, which remains invariant
under $z$-inversion. It follows that the single-particle spectra of $H_{N,\uparrow}^{\infty}$
and $H_{N,\downarrow}^{\infty}$ are both constituted by energies $D_{\infty}\hat\eta_{m\pm}$ ($m=0,1,\ldots$),
where $\hat\eta_{0,\pm}$ are numbers between 0 and $\sqrt\Lambda$, and
\begin{align}
  \hat\eta_{m\pm} = \Lambda^{m+\frac12\mp\frac{\delta}{\pi}}\qquad(m=1,2,\ldots).
  \label{eq:34}
\end{align}
Here again, $\delta$ is a phase shift, given by the expression
\begin{align}
  \label{eq:35}
  \tan\delta =-\pi\dfrac{\ww_{\infty}}{D_{\infty}},
\end{align}

Again we have a line of fixed points. In contrast with the weak-coupling fixed points, however, each
strong-coupling fixed points comprises a single, spin-degenerate band, with the phase shift
$\delta$. This is consistent with the expectation that the magnon  velocity, which controls the difference between the velocities $v\pm u$ in the magnon frame, flows to zero   as $J_\parallel, J_\perp$ flow to strong coupling, as discussed in Sec. \ref{sec:pertRG}.

\subsection{Deviations from the weak-coupling fixed point}
\label{sec:13}

\subsubsection{Deviations}
\label{sec:16}
Before discussing the behavior of the Hamiltonian $\hzero$ in the vicinity of the weak-coupling fixed
point, brief recapitulation of NRG linearization seems appropriate. Given a fixed point $h^{*}$, the
renormalization-group flow of a Hamiltonian $h_{N}$ in the vicinity of $h^{*}$, to linear order in
the deviation between $h_{N}$ and $h^{*}$ is described by the approximate form
\begin{align}
  \label{eq:36}
  \drg{N-1}\,\delta h_{N} = \sum_{p=1,2,\dots} w_{p}\mathcal{O}_{p},
\end{align}
where the coefficients $w_{p}$ depend on $\delta h_{N}$, and the $\mathcal{O}_{p}$ are the
eigenoperators of the renormalization-group transformation $\trg{h_{N}}$, \ie\ each
$\mathcal{O}_{p}$ is an operator with the symmetry of $h_{N}$ satisfying the equality
\begin{align}
  \label{eq:37}
  \trg{\mathcal{O}_{p}} = \lambda_{p}\mathcal{O}_{p},
\end{align}
with eigenvalue $\lambda_{p}$, at the fixed point.

In practice, the eigenoperators are constructed from an infinite sequence of Fermi operators
$\phi_{n}$ ($n=0,1,\ldots$) defined by the equality
\begin{align}
  \label{eq:38}
  \phi_{n} = \sum_{m}\alpha_{nm}\Big[g_{m+}+(-1)^{n}g_{m-}\Big],
\end{align}
where the $g_{m\pm}$ are the operators that diagonalize $H_{\delta}^{*}$, defined by Eq.~\eqref{eq:25}, and
the coefficients $\alpha_{nm}$ are those in Eq.~\eqref{eq:7}; for particle-hole symmetric $h^{*}$, the
$\phi_{n}$ coincide with the $\f{n}$.

Each operator $\mathcal{O}_{p}$ is a linear combination of products of $\phiop{n}$'s and
$\phidop{n}$'s respecting the symmetry of $h^{*}$. The resulting eigenvalue is given by a simple
rule:\cite{Wi75:773,KWW80:1003} each operator $\phiop{n}$ or $\phidop{n}$ with even (odd) index $n$
contributes a factor $\Lambda^{-(n+1)/4}$ ($\Lambda^{-(2n+1)/4}$) to the operators on the \rhs\ of
Eq.~\eqref{eq:36}. The index $n$ of the $\phiop{n}$ operators therefore defines a hierarchy; the
most relevant $\mathcal{O}_{p}$'s are the bilinear forms $\phidop{0}\phiop{0}/\drg{-1}$ compatible with the
symmetry of $h^{*}$, which have eigenvalue $\lambda=1$, next come the bilinear forms
$\drg{-1}(\phidop{0}\phiop{1}+\hc)$, with $\lambda= \Lambda^{-1/2}$, then $\drg{-1}(\phidop{0}\phiop{2}+\hc)$ and
$\drg{-1}(\phidop{1}\phiop{1}+\hc)$, with $\lambda= \Lambda^{-1}$, and so successively.

\subsubsection{Weak-coupling fixed point}
\label{sec:17}
Given the symmetry of the weak-coupling fixed points, which conserve charge and $S_{tot}^{z}$ and
remains invariant under $z$-inversion, its most relevant eigenoperators are
\begin{align}
  \label{eq:39}
  \mathcal{O}^{wc}_{1} &= 
\dfrac1{\drg{N-1}}(\sum_{\mu\nu}\phidop{0\mu}\phiop{0\nu}\pmb\sigma_{\mu\nu}\cdot \mathbf{ S}_{M}+\hc),\\
\label{eq:40}
  \mathcal{O}^{wc}_{2} &=
  \dfrac1{\drg{N-1}}(\phidop{0\uparrow}\phiop{0\uparrow}-\phidop{0\downarrow}\phiop{0\downarrow})S^z_{M},
\end{align}
and
\begin{align}
  \label{eq:41}
    \mathcal{O}^{wc}_{3} &=
\dfrac1{\drg{N-1}}(\phidop{0\uparrow}\phiop{0\uparrow}+\phidop{0\downarrow}\phiop{0\downarrow}).
\end{align}

The operator $\dfrac1{\drg{N-1}}(\phidop{0\uparrow}\phiop{0\downarrow}S^-_{M}+\hc)$, by contrast, is
not an eigenoperator, since it is the linear combination of $\oop{wc}{1}$ e $\oop{wc}{2}$, which
have distinct eigenvalues: as Sec.~\ref{sec:21} will show, the eigenvalue $\lambda_{1}$ of
$\oop{wc}{1}$ depends on the parameters of the Hamiltonian, while $\lambda_{2}=1$.

The deviations of the Hamiltonian $H_{N}$ from $H_{\delta}^{*}$ for small $N$ are therefore
approximately described by the corrections due to the perturbation $\delta H^{wc}_{N}$ defined by
the equality
\begin{align}
  \label{eq:42}
  \delta H^{wc}_{N}= w_{1}\oop{wc}{1}+w_{2}\oop{wc}{2}+w_{3}\oop{wc}{3},
\end{align}
with coefficients $w_{i}$ ($i=1,2,3$) that depend on the coupling constant $J_{\perp}$ and on the
fixed-point phase shifts $\delta_{\pm}$.

\subsubsection{Lanczos operators}
\label{sec:19}

Equations~\eqref{eq:39}-\eqref{eq:41} relate the $\mathcal{O}_{j}$ ($j=1,2,3$) to the Fermi
operators $\phiop{n}$ ($n=0,1,\ldots$). As Eq.~\eqref{eq:38} shows, the latter are linear
combinations of the Lanczos operators $\f{n}$. In particular, as explained in Sec.~\ref{sec:16}, in the
vicinity of particle-hole symmetric fixed points, with phase shifts $\delta=0$ or $\delta=\pi/2$, we
have that $\phiop{n}=\f{n}$ ($n=0,1,2,\ldots$). Since the weak-coupling fixed point is particle-hole
asymmetric, we instead have that
\begin{align}
  \label{eq:51}
  \f{m} = \sum_{n}\gamma_{m,n}\phiop{n},
\end{align}
with coefficients $\gamma_{m,n}$ dependent on the fixed-point phase shifts. For instance,
$\gamma_{0,0}\approx \cos\delta$, while $\gamma_{1,0}\approx\sin\delta$.

It follows from Eq.~\eqref{eq:51} that
\begin{align}
  \label{eq:52}
  \fd{0}\f{0} =& \gamma_{0,0}^{2}\,\phidop{0}\phiop{0} + 
[\gamma_{0,0}\gamma_{0,1}(\phidop{1}\phiop{0}+\hc) &\nr 
&+ \mathrm{other~irrelevant~terms}],&
\end{align}
which shows that $(1/\drg{N-1})\fd{0}\f{0}$ is marginal.

Analogous considerations apply to operators of the form $(1/\drg{N-1})\fd{m}\f{m+1}$
($m=0,1,\ldots$), for which we
have that
\begin{align}
  \label{eq:53}
  \fd{m}\f{m+1} =&\gamma_{m,0}\gamma_{m+1,0}\,\phidop{0}\phiop{0}&\nr
&+(\gamma_{m,0}\gamma_{m+1,1}\,\phidop{0}\phiop{1}
+\gamma_{m,1}\gamma_{m+1,0}\,\phidop{1}\phiop{0}&\nr
&+\mathrm{other~irrelevant~terms}).&
\end{align}

Under particle-hole symmetry, $\gamma_{m,0}$ vanishes unless $m=0$. It follows that the first term on
the \rhs\ of Eq.~\eqref{eq:53} is zero and that $(1/\drg{N-1})\fd{m}\f{m+1}$ is irrelevant in the
vicinity of a particle-hole symmetric fixed point. If the fixed point is particle-hole asymmetric,
by contrast, $\gamma_{m,0}\gamma_{m+1,0}\ne0$ and $(1/\drg{N-1})\fd{m}\f{m+1}$ will be marginal.

\subsection{Weak-coupling fixed point with $u=0$}
\label{sec:18}
For small parameter $u$, an alternative view of the scaled truncated Hamiltonian $H_{N}$ proves
valuable. Recall that this condition is satisfied in the ``slow magnon''  regime $u\ll v$, as  discussed in Sec. \ref{sec:chiralKondo}. Under this condition, we can regard both the terms proportional to $u$ and $J_{\perp}$
on the \rhs\ of Eq.~\eqref{eq:14} as perturbations. The unperturbed Hamiltonian is then given by the
expression
\begin{align}
  \label{eq:54}
  H_{N}^{0} \equiv\dfrac1{\drg{N-1}}
\Bigg\{&\sum_{n=0}^{N-1}t_{n}\left[
v(\fd{n\uparrow}\f{n+1\uparrow}+\fd{n\downarrow}\f{n+1\downarrow})+\hc\right]\nr
&+ 2J_{\parallel}\left(\fd{0\uparrow}\f{0\uparrow}-\fd{0\downarrow}\f{0\downarrow}\right)S^z_{M}&\nr
&+2\vv\left(\fd{0\uparrow}\f{0\uparrow}+\fd{0\downarrow}\f{0\downarrow}\right)
\Bigg\}.
\end{align}

As $N$ grows, $H_{N}^{0}$ approaches a fixed point of the form~\eqref{eq:25}. Consider, now, the
effect of the perturbation
\begin{align}
  \label{eq:55}
  \delta H_{N,u} = -\dfrac{2u}{\drg{N-1}}\sum_{m=0}^{N-1}&
\left[t_{m}(\fd{m\uparrow}\f{m+1\uparrow}-\fd{m\downarrow}\f{m+1\downarrow})S^z_{M}\right.&\nr
&\left.+\hc\right].&
\end{align}

Since $H_{N,0}$ is particle-hole asymmetric, every term on the \rhs\ of Eq.~\eqref{eq:55} is
marginal, as shown by Eq.~\eqref{eq:53}. Given that the coefficients $\gamma_{m,0}$ decay rapidly
with $m$, the contribution of $(\fd{0\uparrow}\f{1\uparrow}-\fd{0\downarrow}\f{1\downarrow})S^z_{M}$
to the matrix elements of $\delta H_{N,u}$ is substantially larger than those of the terms with
$m>0$. To a good approximation, therefore, we can rewrite Eq.~\eqref{eq:55} in the form
\begin{align}
  \label{eq:56}
    \delta H_{N,u} =
    -\dfrac{2ut_{0}}{\drg{N-1}}\left[(\fd{0\uparrow}\f{1\uparrow}
      -\fd{0\downarrow}\f{1\downarrow})S^z_{M}+\hc\right]. 
\end{align}

Substitution of Eq.~\eqref{eq:56} on the \rhs\ of Eq.~\eqref{eq:14} yields the following approximate
expression for the scaled truncated Hamiltonian
\begin{align}
  \label{eq:57}
  \drg{N-1}\tilde H_{N}
=~&v\sum_{n=0}^{N-1}t_{n}
(\fd{n\uparrow}\f{n+1\uparrow}+\fd{n\downarrow}\f{n+1\downarrow}+\hc)&\nr
&-2 u t_{0}\Big(\fd{0\uparrow}\f{1\uparrow}-\fd{0\downarrow}\f{1\downarrow}+\hc\Big)S^z_{M}
\nr
&+ 
J_{\perp}\left(\fd{0\uparrow}\f{0\downarrow}S^-_{M}+\hc\right)&\nr
&+ 2J_{\parallel}\left(\fd{0\uparrow}\f{0\uparrow}-\fd{0\downarrow}\f{0\downarrow}\right)S^z_{M}\nr
&+2\vv\left(\fd{0\uparrow}\f{0\uparrow}+\fd{0\downarrow}\f{0\downarrow}\right).
\end{align}

Section~\ref{sec:15} presents numerical data resulting from the iterative diagonalization of
the Hamiltonian~\eqref{eq:57}. Preliminary to that, however, it seems appropriate to
qualitatively discuss the renormalization-group flow on the basis of perturbative results.

\subsection{Qualitative discussion of the renormalization-group flow}
\label{sec:21}

To visualize the flow of the Hamiltonian~\eqref{eq:57} in renormalization-group space, it is
convenient to split $H_{N}$ into an unperturbed term and a perturbation defined by a linear
combination of the eigenoperators $\mathcal{O}_{n}^{wc}$ ($n=0,1,2$) defined in
Sec.~\ref{sec:13}. Inspection of Eq.~\eqref{eq:57} identifies on the \rhs\ the spin-flip term, that
is, the term proportional to $J_{\perp}$, in the absence of which the Hamiltonian $H_{N}$ can be
easily diagonalized, as Sec.~\ref{sec:4} explained. The spin-flip term can be regarded as
(proportional to) a part of the operator $\oop{wc}{1}$. In order to split the \rhs\ of
Eq.~\eqref{eq:57} into an unperturbed part and a perturbation proportional to $\oop{w}{1}$ we can
choose either
\begin{align}
  \label{eq:46}
  \drg{N-1}\delta H_{N}=~& J_{\parallel}\left[(\fd{0\uparrow}\f{0\downarrow}S^-_{M}+\hc)
\right.&\nr
&\left.+ 2(\fd{0\uparrow}\f{0\uparrow}-\fd{0\downarrow}\f{0\downarrow})S^z_{M}\right]&
\end{align}
or
\begin{align}
  \label{eq:49}
  \drg{N-1}\delta\tilde H_{N}= ~&J_{\parallel}\left[(\fd{0\uparrow}\f{0\downarrow}S^-_{M}+\hc)
\right.&\nr
&\left.-2 (\fd{0\uparrow}\f{0\uparrow}-\fd{0\downarrow}\f{0\downarrow})S^z_{M}\right]&
\end{align}
as the perturbation.

With $\f{0}\to\phiop{0}$, the Hamiltonian in Eq.~\eqref{eq:46} is evidently proportional to
$\oop{wc}{1}$. To show that the Hamiltonian in Eq.~\eqref{eq:49} is also proportional to
$\oop{wc}{1}$ one only has to let $S^{\pm}_{M}\to-S^{\pm}_{M}$, a gauge transformation that switches the
sign of the first term on the \rhs\ without affecting the second. We therefore have two choices. To decide
between them, we have to examine the unperturbed part of the Hamiltonian resulting from each
alternative. Comparison of Eq.~\eqref{eq:57} with Eqs.~\eqref{eq:46}~and \eqref{eq:49} shows that
the coefficient of the term proportional to
$(\fd{0\uparrow}\f{0\uparrow}-\fd{0\downarrow}\f{0\downarrow})S^z_{M}$ in the unperturbed
Hamiltonian stemming from Eq.~\eqref{eq:46} is smaller than the coefficient in the unperturbed
Hamiltonian associated with Eq.~\eqref{eq:49}. It results that $\delta \tilde H_{N}$ is more relevant than
$\delta H_{N}$ and, therefore, that Eq.~\eqref{eq:49} prevails over Eq.~\eqref{eq:46}.

The unperturbed Hamiltonian is, then,
\begin{align}
  \label{eq:79}
  \drg{N-1}\tilde H_{N}^{0} = ~&v\sum_{n=0}^{N-1}t_{n}
(\fd{n\uparrow}\f{n+1\uparrow}+\fd{n\downarrow}\f{n+1\downarrow}+\hc)&\nr
&-2 u\,t_{0}\Big(\fd{0\uparrow}\f{1\uparrow}-\fd{0\downarrow}\f{1\downarrow}+\hc\Big)S^z_{M}\nr
&
+2(J_{\parallel}+J_{\perp})\left(\fd{0\uparrow}\f{0\uparrow}-\fd{0\downarrow}\f{0\downarrow}\right)S^z_{M}&\nr
&+2\vv\left(\fd{0\uparrow}\f{0\uparrow}+\fd{0\downarrow}\f{0\downarrow}\right),
\end{align}
and the perturbation,
\begin{align}
  \label{eq:58}
  \drg{N-1}\delta \tilde H_{N} =  2J_{\perp}\sum_{\mu\nu}\fd{0\mu}\f{0\nu}\,\vec\sigma_{\mu\nu}\cdot\vec S_{ M},
\end{align}
so that $\tilde H_{N}=\tilde H_{N0}+\delta \tilde H_{N}$.

In view of Eqs.~\eqref{eq:39}, after the irrelevant contributions on the \rhs\ of Eq.~\eqref{eq:52}
are disregarded, Eq.~\eqref{eq:58} can be written in the form
\begin{align}
  \label{eq:63}
\drg{N-1}\delta \tilde H_{N} =  2J_{\perp}\gamma_{0,0}^{2}\oop{wc}{1}.
\end{align}

Like the Hamiltonian~\eqref{eq:17}, $\tilde H_{N0}$ is easily diagonalized. Since $S^z_{M}$ is conserved, its eigenvectors
can be classified by the $z$ component of the magnon spin. Given the eigenvalue $s^{z}_{M}$, the \rhs\
of Eq.~\eqref{eq:79} is quadratic and hence numerically diagonalizable. Alternatively, we can
split the \rhs\ of Eq.~\eqref{eq:79} into an $\uparrow$ and a $\downarrow$ terms:
\begin{align}
  \label{eq:59}
\tilde H_{N0} = \tilde H_{N0,\uparrow}+\tilde H_{N0,\downarrow},
\end{align}
where
\begin{align}
  \label{eq:60}
  \tilde H_{N0,\mu}=~&v\sum_{n=0}^{N-1}t_{n}
(\fd{n\mu}\f{n+1\mu}+\hc)&
\nr
&+2\vv\fd{0\mu}\f{0\mu}-u\theta_{z} t_{0}\Big(\fd{0\mu}\f{1\mu}+\hc\Big)\nr
&+(J_{\parallel}+J_{\perp})\theta_{z}\fd{0\mu}\f{0\mu}\qquad(\theta_{z}=\pm1),
\end{align}
and we have introduced the shorthand
\begin{align}
  \label{eq:67}
  \theta_{z} \equiv (2s^{z}_{M})(2\mu),
\end{align}
so that $\theta_{z}=1$ ($\theta_{z}=-1$) when the magnon and holon spins are parallel (antiparallel).

For each $\theta_{z}$, we can now let $\f{n}\to\f{n-1}$ ($n=1,2,\dots, N$) and $\f0\to
c_{0}$ to identify $H_{N0\mu}$ with a resonant-level Hamiltonian, \ie\ a spinless $U=0$ Anderson
Hamiltonian in which $c_{0}$ represents the impurity level and the remaining Lanczos operators
$\f{n}$ represent the holons. 
In this picture, the impurity has energy $2V+\theta_{z}(J_{\parallel}-J_{\perp})$ and its coupling
to the conduction band is $t_{0}(v-u\theta_{z})$. It follows that, for each product $(2s^{z}_{M})(2\mu)=+1$
or $(2s^{z}_{M})(2\mu)=-1$, the conduction levels form a conduction band with uniform phase
shifts given by the expression
\begin{align}
  \label{eq:62}
  \tan\delta(\theta_{z}) = 
  -\pi v\dfrac{V+\theta_{z}(J_{\parallel}+J_{\perp})/2}{(v-\theta_{z}u)^{2}}
  \qquad(\theta_{z}=\pm1).
\end{align}

The weak-coupling fixed point comprises distinct phase shifts. For instance, if the magnon spin is
$\Uparrow$ ($\Downarrow$), the $\uparrow$ holons have the phase shifts in Eq.~\eqref{eq:62} with
$\theta_{z}=1$ ($\theta_{z}=-1$). Equation~\eqref{eq:62} is the analogue of Eq.~\eqref{eq:29}. While
the latter determines the phase shifts of the weak-coupling fixed point for the scaled truncated
Hamiltonian~\eqref{eq:14}, the former determines the phase shifts for the approximate
Hamiltonian~\eqref{eq:57}. With $u=J_{\perp}=0$, the two expressions for the phase shifts
coincide. For small $u/v$ the first-order expansion of Eq.~\eqref{eq:29} reads
\begin{align}
  \label{eq:30}
  \tan\delta_{\pm} = 
  -\pi \dfrac{V\pm J_{\|}/2}{v}\left(1\pm \dfrac uv\right),
\end{align}
while the first-order expansion of Eq.~\eqref{eq:62} for $J_{\perp}=0$ yields
\begin{align}
  \label{eq:31}
\tan\delta(\theta_{z})=  -\pi \dfrac{V+\theta_{z}J_{\|}/2}{v}\left(1+2\theta_{z}\dfrac uv\right).
\end{align}

Compared with Eq.~\eqref{eq:30}, Eq.~\eqref{eq:31} shows that the truncation leading to Eq.~\eqref{eq:57}
leads to renormalization of the anomalous velocity $u$, so that $u\to 2u$. Otherwise, to first
order in $u/v$, the conduction-band phase shifts are identical. We will next show that 
the phase shifts $\delta(\theta_{z})$ control the renormalization-group flow of $H_{N}$ to insure
that the numerical results in Sec.~\ref{sec:15}, computed with the approximate form of the scaled
truncated Hamiltonian in Eq.~\eqref{eq:57}, yield quantitatively reliable conclusions concerning the
magnon-holon interactions.

\subsubsection{Perturbative treatment of the deviations from the weak-coupling fixed point}
\label{sec:22}

The perturbation~\eqref{eq:58} couples states with different $s^{z}_{M}$ and hence breaks the
degeneracy between $s^z_{M}=\Uparrow$, $\mu=\downarrow$ states and $s^z_{M}=\Downarrow$,
$\mu=\uparrow$ states. Consider for example the following degenerate eigenstates of the unperturbed
Hamiltonian:
\begin{align}
  \label{eq:61}
  \ket{\downarrow,\Uparrow} = \gdop{1+\downarrow}\ket{\Omega,\Uparrow},
\end{align}
and
\begin{align}
  \label{eq:64}
  \ket{\uparrow,\Downarrow} = \gdop{1+\uparrow}\ket{\Omega,\Downarrow},
\end{align}
where $\Uparrow$ and $\Downarrow$ indicate the $z$ component of the magnon spin, and
$\gdop{1+\mu}$ creates an holon at the first level above Fermi level with spin component $\mu$.

To compute the first-order correction to the energies of the two states, we have to diagonalize the matrix
\begin{align}
  \label{eq:65}
   & \drg{N-1}\mathcal{\delta H}_{N}^{wc}=&\nr &=2J_{\perp}\gamma_{0,0}^{2}
  \begin{bmatrix}
    \bracket{\downarrow,\Uparrow}{\oop{wc}{1}}{\downarrow,\Uparrow}&
    \bracket{\downarrow,\Uparrow}{\oop{wc}{1}}{\uparrow,\Downarrow}\\
    \bracket{\uparrow,\Downarrow}{\oop{wc}{1}}{\downarrow,\Uparrow}&
    \bracket{\downarrow,\Uparrow}{\oop{wc}{1}}{\downarrow,\Uparrow}
  \end{bmatrix},&
\end{align}
where $\gamma_{0,0}=\cos\delta(-)$, because the holon states $\gdop{1+}$ and the magnon spin are
antiparallel both for $\ket{\downarrow,\Uparrow}$ and $\ket{\uparrow,\Downarrow}$.

The two diagonal elements are identical. Straightforward computation shows that
\begin{align}
  \label{eq:66}
  \bracket{\downarrow,\Uparrow}{\oop{wc}{1}}{\downarrow,\Uparrow}_{N}^{\phantom\dagger}=~&
\bracket{\uparrow,\Downarrow}{\oop{wc}{1}}{\uparrow,\Downarrow}_{N}^{\phantom\dagger}&\nr
=&
-\dfrac{\alpha_{01}^{2}}2\Lambda^{-\frac{N-1}2},&
\end{align}
with the coefficient $\alpha_{01}$ defined by Eq.~\eqref{eq:8}.

The two off-diagonal matrix elements are also identical. Each can be factored into a matrix elements
with holon spin components $\mu=\uparrow$ and another with $\mu=\downarrow$. We for
instance have that
\begin{align}
  \label{eq:47}
  \bracket{\downarrow,\Uparrow}{\oop{wc}{1}}{\uparrow,\Downarrow}_{N}^{\phantom\dagger}=~&
    \bracket{\uparrow,\Downarrow}{\oop{wc}{1}}{\downarrow,\Uparrow}_{N}^{\phantom\dagger}&\nr =~& 
    \bracket{\Omega_{\uparrow},-}{\gop{1+\uparrow}\phidop{0\uparrow}}{\Omega_{\uparrow},+}_{N}^{\phantom\dagger} &\nr
&\times  \bracket{\Omega_{\downarrow},+}{\phiop{0\downarrow}\gdop{1+\downarrow}}{\Omega_{\downarrow},-}_{N}^{\phantom\dagger},&
\end{align}
where $\ket{\Omega_{\mu}, \theta_{z}}$ ($\mu=\uparrow,\downarrow$, $\theta_z=\pm$) denotes the ground
state of the $\mu$ component of the conduction band with phase shift $\delta(\theta_{z})$, given by
Eq.~\eqref{eq:62}.

The two factors on the right-hand side of Eq.~\eqref{eq:47} being equal, we only have to
consider the first one. Since the states $\ket{\Omega_{\uparrow}, +}$ and
$\ket{\Omega_{\uparrow}, -}$ have distinct phase shifts, the matrix element expresses an Anderson orthogonality 
catastrophe and decays with $N$ following the power law\cite{OW81.1553}
\begin{align}
  \label{eq:48}
  \bracket{\Omega_{\uparrow},-}{\gop{1+\uparrow}\phidop{0\uparrow}}{\Omega_{\uparrow},+}_{N}=
\alpha(\bar\delta)\Lambda^{-\frac{N-1}4(1- \bar\delta/\pi)^{2}},
\end{align}
where $\bar\delta=\delta(-)-\delta(+)$, and the $N$-independent coefficient $\alpha(\bar\delta)$
satisfies $\alpha(\bar\delta=0)=\alpha_{01}$; for $\bar\delta\ne0$ the coefficient can
only be computed numerically.

The phase-shift difference $\bar\delta$ depends on the Hamiltonian parameters. From
Eq.~\eqref{eq:62}, it follows that $\bar\delta>0$ for
\begin{align}
  \label{eq:50}
  J_{\perp}+J_{\|} > \dfrac{4uv}{u^{2}+v^{2}} V.
\end{align}

Equations~\eqref{eq:66}~and \eqref{eq:48}, determine the two eigenvalues of the perturbative matrix~\eqref{eq:65}:
\begin{align}
  \label{eq:43}
  \delta\varepsilon_{\pm}= -\dfrac{J_{\perp}}{\drg{1}}\cos^{2}\delta(-)\left[\alpha_{01}^{2}\pm
  \alpha^{2}(\bar\delta)\Lambda^{\frac{N-1}4\frac{\bar\delta}{\pi}  (2-\frac{\bar\delta}{\pi})}\right],
\end{align}
where the eigenvalues $\delta\epsilon_{+}$ and  $\delta\epsilon_{-}$ correspond to the triplet and
singlet combinations of $\ket{\downarrow,\Uparrow}$ and $\ket{\uparrow,\Downarrow}$, respectively.

If the parameters are such that inequality~\eqref{eq:50} is satisfied, $\bar\delta$ will be
positive.  Equation~\eqref{eq:43} then shows that the splitting between the triplet and the singlet
will grow as a small positive power of the inverse energy scale $1/\drg{N-1}$. The operator $\oop{1}{wc}$ is
therefore relevant and will drive the scaled truncated Hamiltonian away from the weak-coupling fixed
point.

The renormalization-group evolution expressed by Eq.~\eqref{eq:43} differs only quantitatively from
the evolution for the standard Kondo model. We therefore expect the Hamiltonian to cross over from
the weak-coupling to the strong-coupling fixed points under the influence of
$\oop{wc}{1}$. Nonetheless, since perturbation theory is only reliable in the vicinity of the fixed
point and hence inadequate to describe the crossover, numerical treatment becomes necessary.

\subsection{Numerical results}
\label{sec:15}

The numerical diagonalization of the approximate expression for the scaled truncated Hamiltonian
$H_{N}$ in Eq.~\eqref{eq:57} follows the procedure outlined in Sec.~\ref{sec:7}. We choose the
discretization parameter $\Lambda=3$ and start out with $N=1$, i.~e., with the basis of 32 many-body
states defined by the magnon-spin operator $S^z_{M}$ and the Lanczos operators $f_{0}$ and
$f_{1}$. The conservations of charge and $z$-component of the total spin, and the invariance of the
Hamiltonian under $z\leftrightarrow-z$ inversion reduce the projection of $H_{1}$ onto this basis to
a block-diagonal form that yields to easy numerical diagonalization. The resulting eigenvalues and
eigenvectors seed the iterative cycle, which is stopped at $N=49$, when $\drg{N-1}$ becomes smaller
than $10^{-10}D$. 

Ultraviolet truncation starts at iteration $N=4$, when the highest scaled energies first exceed the
ultraviolet cutoff $E_{uv}=20$ and are discarded. Our discussion being focused on
renormalization-group flows, not on the computation of physical properties, the infrared cutoff
$\epsilon_{IR}^{\phantom{\dagger}}$ needs not be defined. The iterative procedure is very efficient:
with $\Lambda=3$ and $E_{uv}=20$, in a standard desktop computer a complete run takes times of the
order of 100 seconds.

\subsubsection{Flow in renormalization-group space}
\label{sec:12}

To describe the trajectory of the scaled truncated Hamiltonian in renormalization-group space, this
section presents numerical data for the $N$ dependence of illustrative scaled many-body energies
resulting from the iterative diagonalization of $H_{N}$ for various choices of the model parameters
$u$, $J_{\perp}$, $J_{\parallel}$, and $\vv$. All parameters are expressed in units of $v$. Given
that $\trg{H_{N}}$ transforms $H_{N}$ into $H_{N+2}$, we plot the energies as a function of even
iteration number $N$. With $N=\mathrm{even}$ the spectrum of the strong-coupling fixed point is
split into $N/2$ positive and $N/2$ negative single-particle energies, which makes deviations from
particle-hole symmetry more visible than with $N=\mathrm{odd}$.

\begin{figure}[!th]
\centering
  \includegraphics[width=0.95\columnwidth]{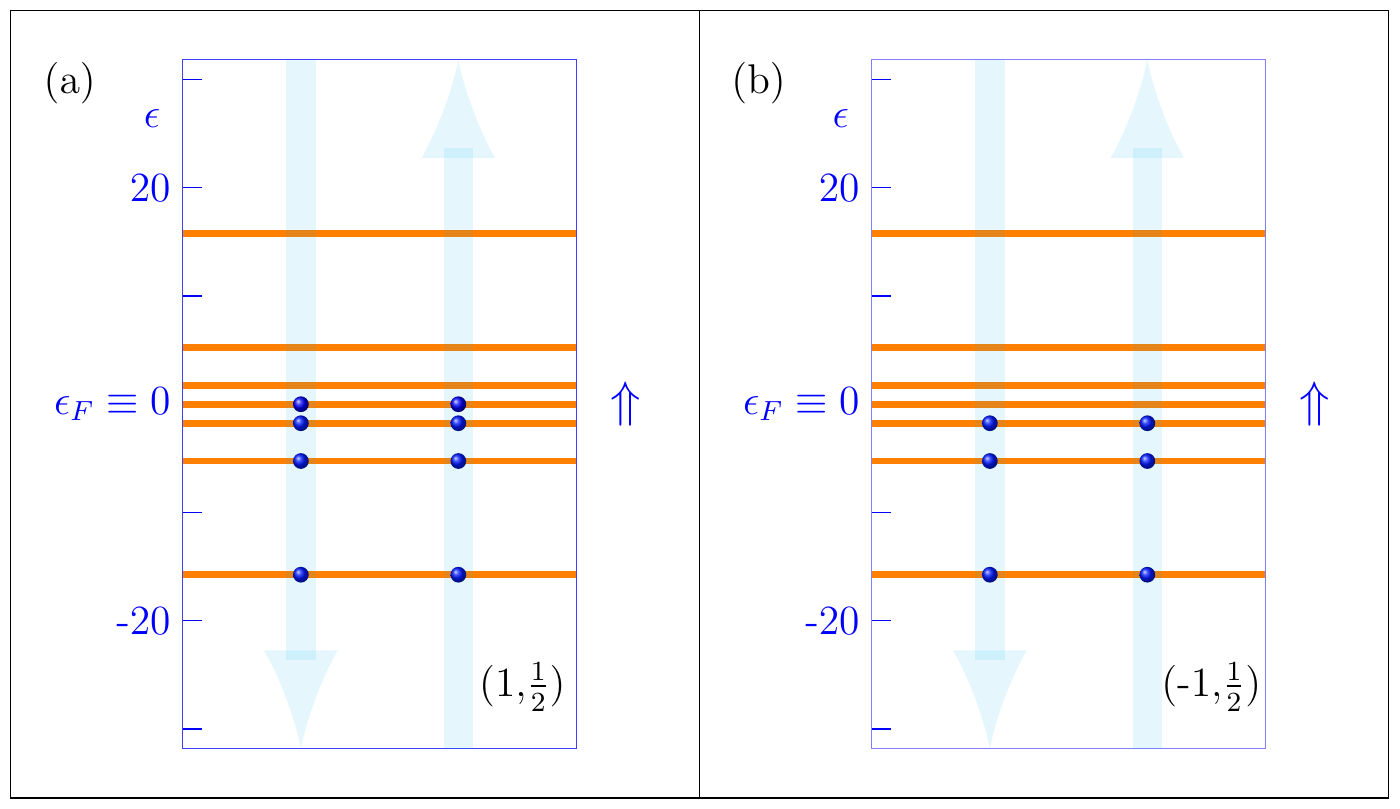}
  \caption{ (color online) Lowest-energy eigenstates in two sectors of the even-$N$ particle-hole symmetric
  weak-coupling fixed-point Hamiltonian with $J_{\|}=u=\vv=0$. Panels (a) and (b) show the
  lowest many-body eigenstate in the $(q=1, s=1/2)$ and $(-1,1/2)$ sectors, respectively. The
  horizontal lines represent the single-particle energies resulting from the diagonalization of
  $H_N$, the blue spheres mark the filled levels, and the symbol $\Uparrow$ indicates the $z$
  component of the magnon spin. Measured from the ground state, the many-body energies of both
  states are zero. \label{fig:00}}
\end{figure}

\begin{figure}
  \centering
  \includegraphics[width=0.95\columnwidth]{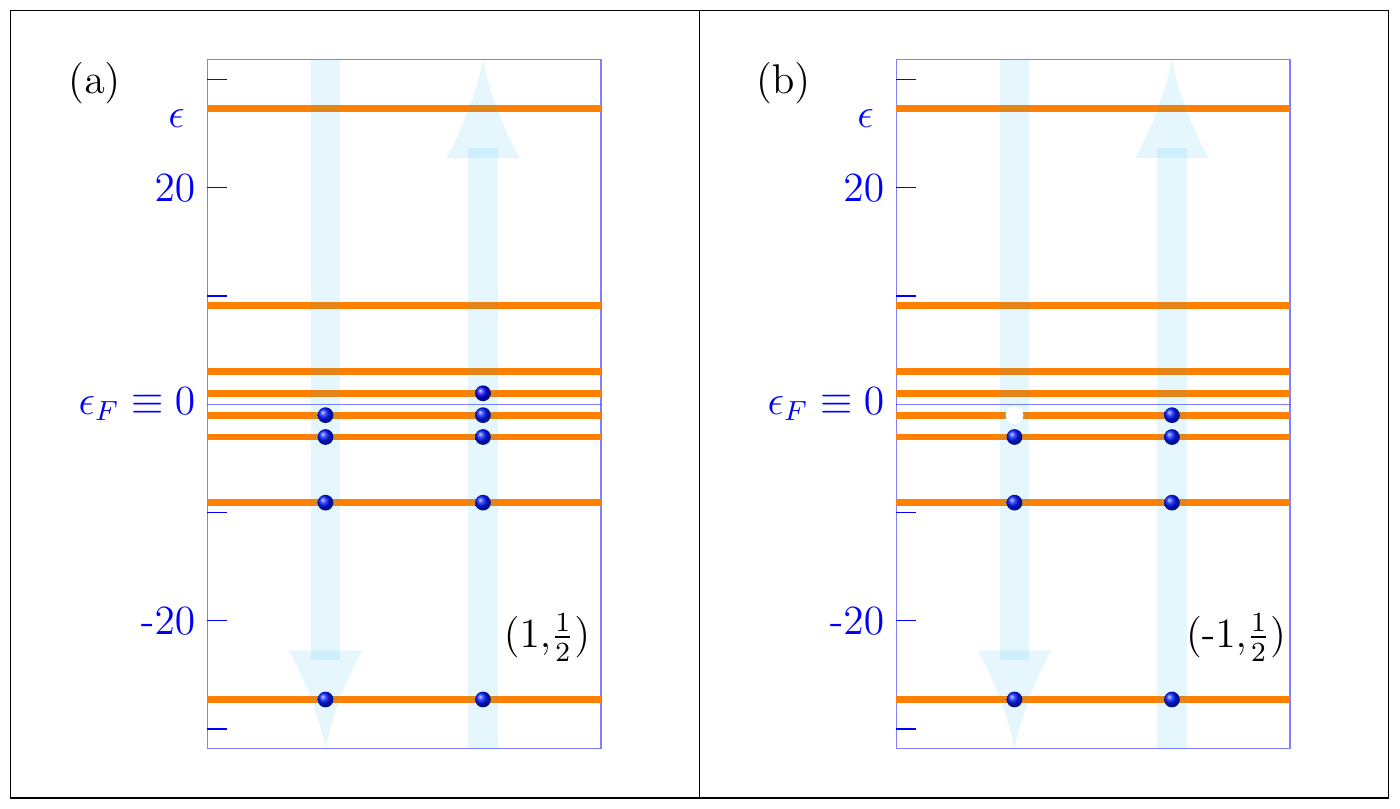}
  \caption{ (color online) Lowest-energy eigenstates in two sectors of the even-$N$ particle-hole symmetric
    strong-coupling fixed-point Hamiltonian with $u=\vv=0$. Panel (a) shows the lowest many-body
    eigenstate at the $(q=1,s=1/2)$ sector, and panel (b) shows the lowest many-body eigenstate of
    the $(-1,1/2)$ sector. The horizontal lines are the single-particle energies resulting from the
    diagonalization of $H_N$, and the blue spheres indicate the filled levels. Measured from the
    ground state, in which all negative levels are filled, both states have scaled energy
    $\eta_1^*\approx0.8$, for discretization parameter $\Lambda=3$.}
  \label{fig:0}
\end{figure}

For clarity, for each run we show the $N$ dependence of the lowest scaled energies $E_{N}$ in the
$(q=1,s_{z}=1/2)$ and $(-1,1/2)$ sectors of the Fock space, where the charge $q$ is measured from the
charge of the half-filled band, the energies are measured from the ground state, and $s_{z}$ denotes
the $z$ component of the total spin. At the weak-coupling and strong-coupling fixed points, the two
states correspond to simple holon configurations, as illustrated by Figs.~\ref{fig:00}~and
\ref{fig:0}. 

Figure~\ref{fig:00} shows the two many-body states at the weak-coupling particle-hole symmetric fixed
point. Since the number of single-particle states is $N+1$, one of the levels lies at $\epsilon=0$
and has ground-state occupation $n_{0}=1$. A second holon (no holon) occupies the $\epsilon=0$
level in panel (a) [panel (b)], which represents the lowest-energy state in the $(1,1/2)$
[$(-1,1/2)$] sector. The two states are degenerate with the ground state.

Figure~\ref{fig:0} shows the two states at the corresponding strong-coupling fixed point. The magnon
spin now being locked into a singlet with a holon, the remaining $N$ single-particle levels are
symmetrically distributed around the Fermi level. The minimum-energy state in the $(1,1/2)$ [$(-1,1/2)$]
sector contains a single particle (hole) at the lowest (highest) level above (below) the Fermi
level. The two states are degenerate, with scaled energy $\hat\eta_{1+}=\hat\eta_{1-}$ relative to
the ground-state energy. The energies $\hat\eta_{1\pm}$ are approximately described by
Eq.~\eqref{eq:26}, with $\delta=\pm\pi/2$. For $\Lambda=3$, in particular, $\hat\eta_{1\pm}=0.8$.

The following sections describe the $N$ dependence of the lowest many-body energies in the
$(\pm,1/2$) sectors for various model parameters. For subsequent reference, we start out with an
example of particle-hole symmetry.

\subsubsection{Particle-hole symmetric Hamiltonians ($u=\vv=0$)}
\label{sec:14}
\begin{figure}[!th]
  \centering
  \includegraphics[width=0.95\columnwidth]{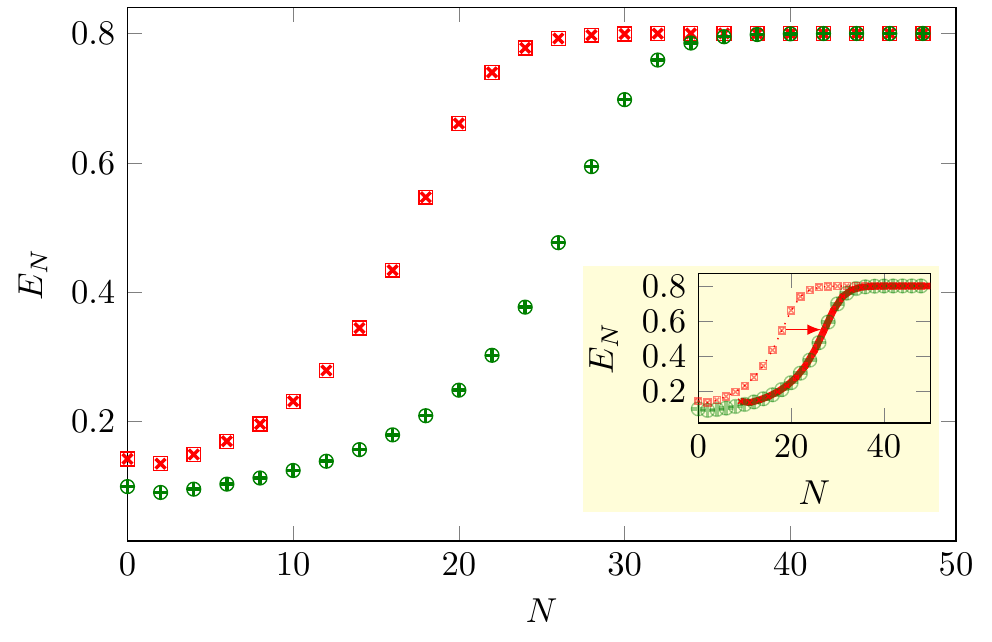}
  \caption{(color online) Scaled lowest energies $E_N$ as functions of iteration number $N$ for
    sectors $(1,1/2)$ (open circles) and $(-1,1/2)$ (crosses). The green circles and crosses
    represent eigenvalues of the Hamiltonian~\eqref{eq:57} with $u=\vv=0$, $J_{\perp}=0.2\,v$, and
    $J_{\parallel}=0.01\,v$. As $N$ grows, the renormalization-group transformation $\trg{H_{N}}$
    drives the Hamiltonian from the vicinity of the weak-coupling fixed point, where $E_{N}^{*}=0$,
    to the strong-coupling fixed point, where $E_{N}^{*}=0.8$. For comparison, the evolution of the
    eigenvalues for a symmetric Hamiltonian, with $J_{\parallel}=J_{\perp}=0.2\,v$, is also
    shown. The inset shows that, horizontally shifted by $\Delta{N}=9.1$, the isotropic curve
    (red solid line) coincides with the anisotropic curve (green crosses and open circles).}
  \label{fig:1}
\end{figure}

With $u=\vv=0$, Equation~\eqref{eq:1} describes a particle-hole symmetric, anisotropic Kondo
Hamiltonian, the renormalization-group trajectory of which has long been known to merge onto that of
an isotropic Hamiltonian.\cite{TW83:453} We therefore expect $H_{N}$ to flow from the vicinity of
the particle-hole weak-coupling fixed point to the strong-coupling fixed point. Figure~\ref{fig:1}
shows an example, with $J_{\perp}=0.2v$ and $J_{\parallel}=0.01v$. The particle-like eigenstates in
the $(1,1/2)$ sector are degenerate with the hole-like eigenstates in the $(-1,1/2)$ sector. The
open circles depicting the lowest-energy eigenvalue of $H_{N}$ in the $(1,1/2)$ sector therefore
coincide with the $+$ signs depicting the lowest-energy eigenvalue in the $(-1,1/2)$ sector as the
Hamiltonian crosses over from near the weak-coupling fixed point to the strong-coupling fixed
point. For comparison, the crossover for the isotropic model with $J_{\perp}=J_{\parallel}=0.2\,v$
is also shown. The inset shows that the crossover is universal, a horizontal shift $\Delta N= -9.1$
of the anisotropic curve being sufficient to make the two plots coincide.

\subsubsection{Particle-hole asymmetry}
\label{sec:20}

\begin{figure}
  \centering
  \includegraphics[width=0.95\columnwidth]{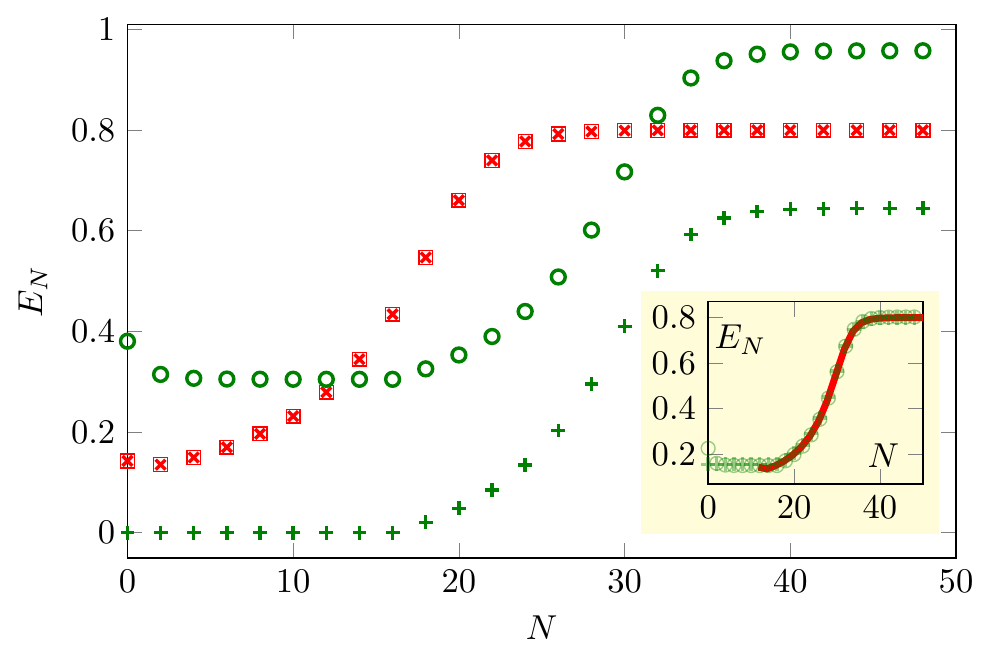}
  \caption{(color online) Scaled lowest energies in sectors $(\pm1,1/2)$ as a function of iteration
    number for the indicated Hamiltonian parameters. As in Fig.~\ref{fig:1}, the corresponding
    energies for the isotropic, particle-hole symmetric Hamiltonian are shown for comparison.  The
    symbol convention follows Fig.~\ref{fig:1}. The scattering potential $\vv$ breaks particle-hole
    symmetry and shifts the $(1,1/2)$ [$(-1,1/2)$] energies up (down). The crossover between the
    weak-coupling and the strong-coupling fixed points nonetheless follows the universal behavior in
    Fig.~\ref{fig:1}, as evidenced by the inset, which show the particle-hole symmetric, isotropic
    curve (red, solid line), horizontally displaced by $\Delta N=11.7$, and the asymmetric curves
    (green open circles and crosses) shifted vertically to make the large $N$ energies match the
    particle-hole symmetric fixed-point energies.}
  \label{fig:2}
\end{figure}

Figure~\ref{fig:2} shows the $N$ dependence of the minimum energies in the $(1,1/2)$ and $(-1,1/2)$
sectors for $J_{\perp}=0.2\,v$, $J_{\parallel}=0.01\,v$, and $\vv=0.10\,v$, with $u=0$. The term
proportional to $\vv$ on the \rhs\ of Eq.~\eqref{eq:57} breaks the degeneracy between the $(1,1/2)$
and $(-1,1/2)$ energies and shifts the phases of the fixed-point single-particle levels, as
discussed in Section~\ref{sec:9}. The phase shifts at the weak-coupling fixed point affect the
renormalization-group flow and delay the crossover to the strong-coupling fixed point. Universality is
nonetheless preserved, as indicated by the congruence in the inset, which compares the isotropic
curve in Fig.~\ref{fig:1} with the two particle-hole asymmetric curves, both shifted horizontally by
$\Delta N=-11.7$ and displaced vertically to insure agreement at large $N$.

\begin{figure}
  \centering
    \includegraphics[width=0.95\columnwidth]{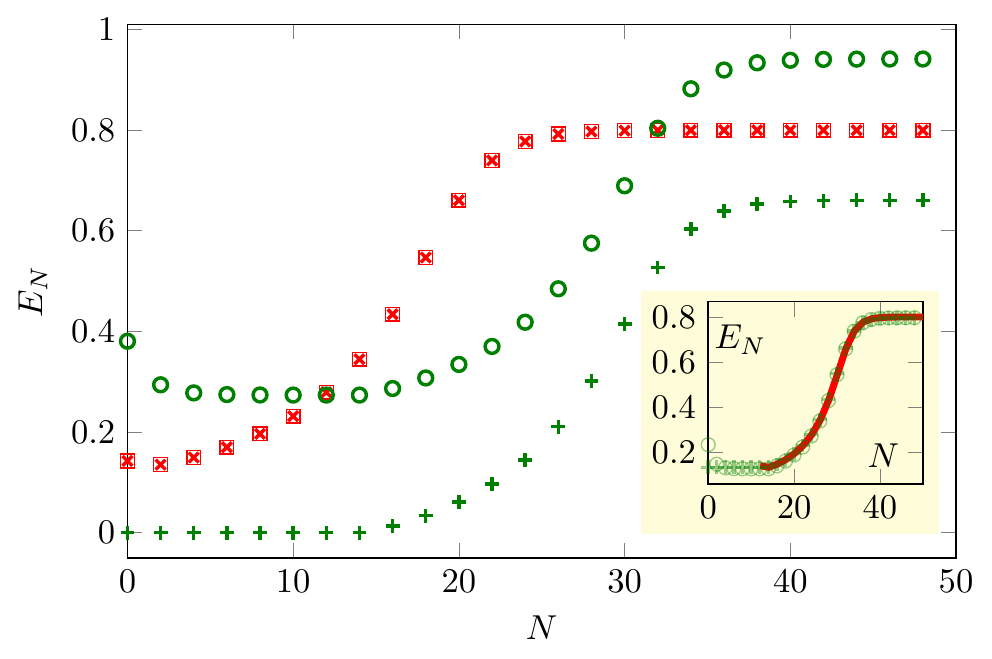}
    \caption{ (color online) Scaled minimum energies in the sectors $(\pm1,1/2)$ as functions of the iteration
      number for the indicated Hamiltonian parameters. We follow the symbol convention in
      Fig.~\ref{fig:1} and for comparison reproduce the isotropic curve in Fig.~\ref{fig:1}. As in
      Fig.~\ref{fig:2}, the energies in the $(1,1/2)$ [$(-1,1/2)$] sector are shifted up (down) relative to the
      particle-hole symmetric energies. To show that universality is preserved, the inset shows the
      particle-hole symmetric curve (red, solid line) displaced horizontally by $\Delta N=12$,
      superimposed upon the particle-hole asymmetric curves (green, open circles and crosses)
      displaced vertically to insure agreement at the large $N$ limit.}
    \label{fig:3}
  \end{figure}

Figure~\ref{fig:3} shows results for $u\ne0$. The minimum energies in the $(\pm1,1/2)$ sectors for
$J_{\perp}=0.2\,v$, $J_{\parallel}=0.2\,v$, and $\vv=0.10\,v$, with $u=0.2\,v$ are plotted as
functions of the iteration number $N$. Comparison with the flow in Fig.~\ref{fig:2} show that the
spin-dependent contribution to the bandwibandwidthdth, \ie\ the term proportional to $u$ on the right-hand
side of Eq.~\eqref{eq:57} perturbs two features of the renormalization-group flow: the weak-coupling
(strong-coupling) phase shifts $\delta_{wc}$ ($\delta_{sc}$) is closer to its isotropic value
$\delta_{wc}=0$ ($\delta_{sc}=\pi/2$), and the crossover from the weak-coupling to the
strong-coupling fixed point is shifted to the right. The delayed crossover indicates that   the
operator $\mathcal{O}_{1}^{wc}$, defined in Sec.~\ref{sec:13}, is less relevant. This conclusion
agrees with Eq.~\eqref{eq:62}, which shows that $u$ reduces $\delta(-)$ and increases $\delta(+)$
and therefore diminishes the difference $\bar\delta$, which controls the exponent on the \rhs\ of
Eq.~\eqref{eq:43}. The $u$-dependent ground-state phase shifts and the universal crossover confirm
that the term proportional to $u$ in the Hamiltonian is a marginal operator, which displaces the
high-energy and the ground-state Hamiltonians along the lines of weak- and strong-coupling fixed
points, respectively.

\begin{figure}[!th]
  \centering
  \includegraphics[width=0.95\columnwidth]{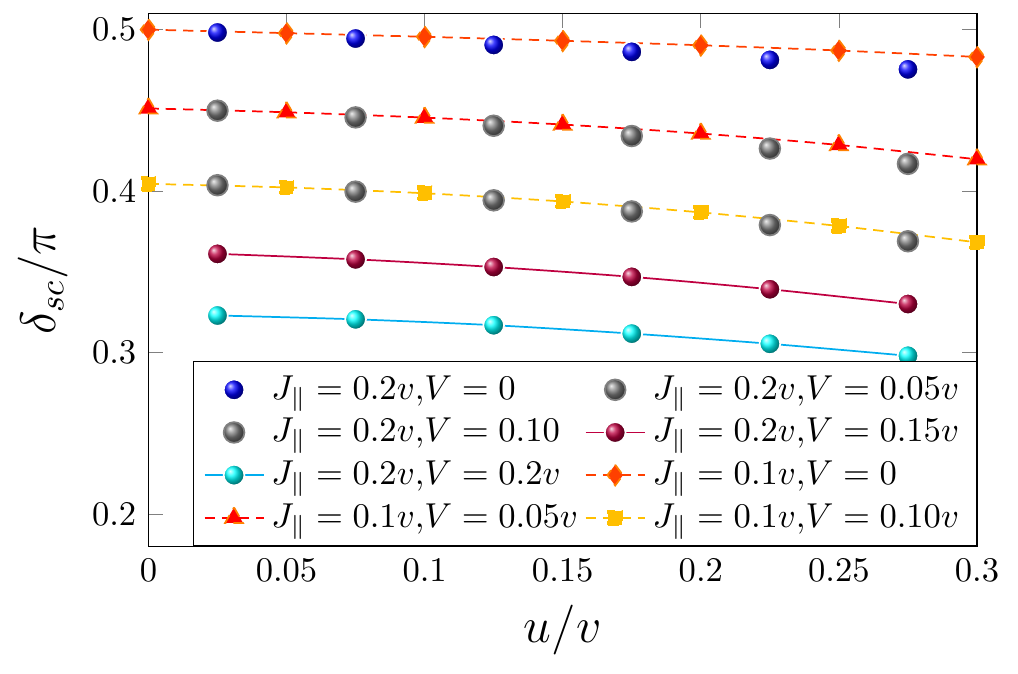}
  \caption{ (color online) Strong-coupling fixed-point phase shifts as a function of the anomalous velocity $u$ at
    fixed $J_{\perp}=0.2\,v$, $J_{\parallel}=0.1\,v$, and $\vv=0.1\,v$. The phase shifts are obtained
    from Eq.~\eqref{eq:34}, the single-particle energies $\eta_{m\pm}$ being identified as the limit
    to which the eigenvalue of the many-body state $\gdop{m}$ ($m=2$) \textemdash sector $(1,1/2)$
    tends as $N$ grows past the crossover. The isotropic limit of the phase shift, not shown, is
    $\delta/\pi=0.5$.}
  \label{fig:4}
\end{figure}

  Analogous results are obtained for other parametrical choices. In particular, as $u$ grows at
  fixed $J_{\perp}$, $J_{\parallel}$, and $\vv$, the crossover is displaced to progressively higher
  $N$, while the strong-coupling phase shifts $\delta_{sc}$ are slightly displaced to higher or lower values,
  chiefly depending on $\vv$. Figure~\ref{fig:4} shows $\delta_{sc}$ as a function of the ratio
  $u/v$ for fixed $J_{\perp}=0.2v$ and illustrative choices of $J_{\|}$ and $\vv$. For $u=0$, we
  expect the weak-coupling phase shifts $\delta(\theta_{z})$ to be given by Eq.~\eqref{eq:43}, and from the Friedel sum
  rule, expect the strong-coupling phase shift to differ from $\delta(\theta_{z})$ by $\pi/2$. We
  therefore expect $\delta_{sc}$ to obey the relation
  \begin{align}
    \label{eq:68}
    \cot\delta_{sc}=-\pi\dfrac{V}{v},
  \end{align}
  which is approximately satisfied for all data in Fig.~\ref{fig:4}. 

  As $u$ grows, $\delta_{sc}$ diminishes for small $\vv$ and grows as $\vv$ exceeds $0.05v$. In all
  cases, however, $\delta_{sc}$ depends weakly on $u$ and is nearly independent of either
  $J_{\perp}$ or $J_{\|}$. Roughly speaking, therefore, the strong-coupling phase shifts are
  approximately described by Eq.~\eqref{eq:68} over the parametrical space of the Hamiltonian.
  Although the approximation distinguishing Eq.~\eqref{eq:57} from the scaled truncated
  Hamiltonian~\eqref{eq:14} restrict our analysis to small $u$, these results ratify the conclusion
  that the anomalous velocity is a marginal operator whose effect upon the physical properties of
  the system under study is limited to reducing the Kondo temperature and (slightly) displacing the
  low-temperature Hamiltonian along the line of strong-coupling fixed points.

\section{Conclusions}
\label{sec:summary}
In this work, we considered the physics of doping a one-dimensional
ferromagnetic insulator. Assuming spin-charge separation in the
strongly interacting regime, we modelled the charge excitations  by spinless
fermions (holons) and the spin excitations by magnons. 

The interaction between magnons and gapless holons in the limit  of small  doping gives rise to infrared singularities in perturbation theory. The  mechanism behind these singularities is clarified by mapping to an 
 effective field theory  equivalent to an anisotropic Kondo model.  In our case,  the holons play the role of the conduction electrons and a magnon with momentum commensurate with the holon Fermi momentum $p_F$ plays the role of a mobile impurity. The internal degree of freedom of the impurity is a chirality pseudospin, which  indicates whether  the momentum   is closest to $p_F$
or $-p_F$.  A pseudospin flip corresponds to a $2p_F$ scattering
process. Another difference from the conventional  Kondo model  is that the impurity (magnon) is
mobile rather than localized. By applying a Galilean transformation to the frame moving with the magnon velocity $u$, we find that the effective   Hamiltonian contains  two holon bands with       velocities $v+u$ or $v-u$ depending on whether the holons move in the same or opposite direction as the magnon. 

We  investigated the effects of the magnon-holon interaction in the nonperturbative,  low-energy regime using the numerical renormalization group. We  generalized the numerical method to treat the effects of a new marginal perturbation proportional to the magnon velocity $u$. The renormalization group flow takes the Hamiltonian  from a line of weak-coupling fixed points characterized by two distinct  phase shifts (related to scattering between the moving magnon and the  two holon bands) to another line of strong-coupling fixed points with a single phase shift $\delta_{sc}$. The existence  of   only one phase shift can be interpreted in terms of the vanishing group velocity of the single-magnon excitation when the magnon momentum approaches $ p_F$. By analogy with the original Kondo effect, we expect that at the strong-coupling fixed point the magnon pairs up with a holon to form a  singlet of the chirality pseudospin. On the other hand, we should  stress that, since the total magnetization is conserved, the real spin carried by the magnon is not screened. The residual potential scattering between the remaining holons and the magnon-holon pseudospin singlet accounts for the nonuniversal phase shift at the strong-coupling fixed point. Finally, we showed that  the initial  value of  $u$ at the weak-coupling fixed point  has a similar  effect to the potential scattering term, namely it drives the phase shift away from the particle-hole symmetric value $\delta_{sc}=\pi/2$.

The results of this work suggest that it would be interesting to study dynamical response functions probing the spin excitation spectrum of one-dimensional itinerant ferromagnets when the magnon momentum approaches the holon Fermi momentum. In particular,   the  single phase shift $\delta_{sc}$   should have consequences for the power-law singularity  at the single-magnon threshold in the transverse spin structure factor, which is equivalent to the magnon spectral function.\cite{Kamenev:2009ez} Another interesting question is whether the Kondo effect discussed here can drive a nontrivial  instability of the metallic ferromagnetic phase of a lattice model such as   Tasaki's model in Eq. (\ref{tasakiHamiltonian}), since it is expected to lower the energy of the single-magnon excitation.  The behavior of the magnon spectral function  in  lattice models, such as  Tasaki's model in Eq. (\ref{tasakiHamiltonian}), could be investigated numerically using time-dependent density matrix renormalization group methods.\cite{White:PhysRevLett.93.076401,Schollwock201196}

\begin{acknowledgments}
This work was supported by the FAPESP (H.P.) and CNPq (R.G.P. and L.N.O.).
\end{acknowledgments}

\bibliography{magnonkondonrg.bbl}

\end{document}